%% file: cdbd_arxiv_sub.tex
\documentclass[useAMS,usenatbib]{mn2e}

\usepackage{amsfonts}
\usepackage{amssymb}
\usepackage{graphicx}
\usepackage{amsmath}

\include{aasmacros}

\title[Direct Shear Mapping]{Direct Shear Mapping - a new weak lensing tool}
\author[C. O. de Burgh-Day, E. N. Taylor, R. L. Webster and A. M. Hopkins]{C. O. de Burgh-Day$^{1,2,3}$\thanks{cdbd@student.unimelb.edu.au$\,$(CDBD); ent@ph.unimelb.edu.au$\,$(ENT); r.webster@unimelb.edu.au$\,$(RLW); ahopkins@aao.gov.au$\,$(AMH)}, E. N. Taylor$^{1,3}$\footnotemark[1], R. L. Webster$^{1,3}$\footnotemark[1] and A. M. Hopkins$^{2,3}$\footnotemark[1]\\
$^{1}$School of Physics, David Caro Building, The University of Melbourne, Parkville VIC 3010, Australia\\
$^{2}$The Australian Astronomical Observatory, PO Box 915, North Ryde NSW 1670, Australia\\
$^{3}$ARC Centre of Excellence for All-sky Astrophysics (CAASTRO) }

\begin{document}

\date{Accepted yyyy Month dd. Received yyyy Month dd; in original form yyyy Month dd}

\pagerange{\pageref{firstpage}--\pageref{lastpage}} \pubyear{????}

\maketitle

\label{firstpage}

\begin{abstract}	
We have developed a new technique called Direct Shear Mapping (DSM) to measure gravitational lensing shear directly from observations of a single background source. The technique assumes the velocity map of an un-lensed, stably-rotating galaxy will be rotationally symmetric. Lensing distorts the velocity map making it asymmetric. The degree of lensing can be inferred by determining the transformation required to restore axisymmetry.
This technique is in contrast to traditional weak lensing methods, which require averaging an ensemble of background galaxy ellipticity measurements, to obtain a single shear measurement. We have tested the efficacy of our fitting algorithm with a suite of systematic tests on simulated data. 
We demonstrate that we are in principle able to measure shears as small as 0.01. In practice, we have fitted for the shear in very low redshift (and hence un-lensed) velocity maps, and have obtained null result with an error of $\pm 0.01$. This high sensitivity results from analysing spatially resolved spectroscopic images (i.e.  3D data cubes), including not just shape information (as in traditional weak lensing measurements) but velocity information as well. Spirals and rotating ellipticals are ideal targets for this new technique. Data from any large IFU or radio telescope is suitable, or indeed any instrument with spatially resolved spectroscopy such as SAMI, ALMA, HETDEX and SKA. 

\end{abstract}



\bibliographystyle{mn2e}
\section{Introduction}

Weak gravitational lensing maps matter distributions in the universe, both baryonic and dark (e.g. \citet{1993ApJ...404..441K}). This paper explores the enhanced potential of weak lensing in three dimensions \citep{2006ApJ...650L..21M,2002ApJ...570L..51B}. We describe a methodology to obtain a shear measurement from a 
3D data cube of a single weakly lensed galaxy. This technique will allow us to measure the size and shape of dark matter distributions around individual galaxies at low redshifts. 

\noindent Conventional 2D weak lensing techniques rely on measuring the (two-dimensional) shapes of many ($\gtrsim 100$) images in a field. With the assumption that the images in the field should have no preferred orientation, one can infer the presence of a shear field if any correlation in alignments is detected. 
There are a number of statistical approaches \citep{2010MNRAS.405.2044B,2006MNRAS.368.1323H,1995ApJ...449..460K,2003MNRAS.338...48R} used to perform this analysis. The statistical uncertainty of a particular weak lensing survey depends on the total survey area and the number density of perfectly measured galaxies. The dominant source of uncertainty in all of these methods is shape noise: the error in the measurement due to the intrinsic and random orientations of the images in the sample. Using these techniques, one would typically need $\sim 10$ objects to measure a shear of $10\%$ on arcminute scales.

\noindent A powerful enhancement of the weak lensing method was proposed by \citet{2002ApJ...570L..51B}, followed by \citet{2006ApJ...650L..21M}. Theoretically, the rotation curves of regularly rotating elliptical and spiral galaxies will have maximum and minimum values in the projected velocity maps. These coincide with the major and minor axes of the projected 2D image. In gravitational lensing, the Equivalence Principle requires that photons of different energies are affected similarly. Thus weak lensing will shear the velocity maps and distort the 2D image, causing the angle between maximum and minimum rotation axes to deviate from $90^{\circ}$. While there are other kinematic effects which also introduce perturbations into the velocity map of a galaxy, weak gravitational lensing has a unique signature. An un-sheared velocity map is symmetrical about the major and minor axis, while a sheared velocity map loses these symmetries. Since we can predict the properties expected in an un-sheared velocity map (namely that the angle between the major and minor axis is orthogonal, or put another way, that the velocity is symmetrical about these axes), a metric which measures the presence and strength of the shear field between the observer and the background galaxy can be designed.

\noindent \citet{2002ApJ...570L..51B} suggested using the distortion in the rotation curve to measure shear. He fitted for shear in an inclined ring model using a Monte-Carlo routine, concluding that while currently a typical shear could not be measured by this method, it would be possible to measure it with future higher resolution surveys.
\citet{2006ApJ...650L..21M} suggested a similar method, but fitted for the entire velocity map. This has the advantage of utilising additional information from the velocity map, and potentially avoiding problems the concentric-ring method would encounter when fitting for shear in warped or disturbed disks (since in that case each concentric ring will recover a different shear). Morales' method involved measuring the angle between the major and minor rotation axes, and obtaining a measure of the shear strength from the deviation of the fitted angle from orthogonal.

\noindent We have extended Morales' work, and developed a technique to measure the shear vector directly from the 3D data cube of a single weakly lensed galaxy. Our technique is called Direct Shear Mapping (DSM) and utilises a Monte-Carlo Markov Chain (MCMC) algorithm to search for asymmetries in the 3D data. The MCMC function used is called \textsc{emcee}\footnote{the Emcee Hammer, http://dan.iel.fm/emcee/} \citep{2013PASP..125..306F}.  We have performed a suite of  tests on simulated data to characterise the efficiency and accuracy of the fitting algorithm, and to understand the limits of its fitting range. Additionally we have tested the technique on un-sheared data to ensure we can recover a null result at low redshift, and determine realistic systematic errors.

\noindent  Using a symmetry search method rather than focusing on the velocity axes directly allows us to use all the available data, giving a statistically better fit. This method is unique among weak lensing measurement methods in that it uses the velocity map of an object to measure the shear field, and does not rely on fitting for the shape of the galaxy. Rather than obtaining a single measure of the shear over a wide area of sky, we can determine a shear value for a single background galaxy. Hence we are able to measure the mass of an individual foreground dark matter halo.

\noindent We have demonstrated that we can measure shears as small as $\gamma \sim 0.01$ in realistic simulations and expect to extend this to observational datasets (Taylor et al, in prep.). Two-dimensional weak lensing methods are able to statistically measure the mass and structure of an individual dark matter halo and establish its relationship to other observables, for example baryonic mass (eg \citet{2014MNRAS.437.2111V}).
It may also be possible to probe for significant substructure in the shear field on small angular scales by making a number of independent shear measurements around a dark matter halo, a problem which shape-fitting and moment-measurement methods cannot address, since they are blind to variations on smaller scales. We already have a first probable detection (Taylor et al, in prep.), and there are a number of existing and upcoming instruments and surveys which will produce data products well suited to our technique, including observations with SPIRAL \citep{2001PASP..113..215K}, SAMI \citep{2012MNRAS.421..872C} and CALIFA \citep{2012A&A...538A...8S} in the optical, and ASKAP and the SKA at radio frequencies \citep{2008ExA....22..151J,2000pras.conf..203S,2007astro.ph..3746B}. \\

\noindent This paper is organised as follows. In section~\ref{sec:Weaklensing} we briefly review weak lensing theory. In section~\ref{sec:Methods} we describe DSM, and the structure of the DSM fitting algorithm. In section~\ref{sec:Modellingandtestingofmethod} we describe the generation of synthetic galaxy data, and present the results of a suite of systematic tests of the DSM method. In section~\ref{sec:Applicationtorealdata} we present fits to low-redshift data from the literature, demonstrating the ability of DSM to recover a null result. Conclusions are presented in section~\ref{sec:Conclusions}.

\section{Weak lensing}
\label{sec:Weaklensing}
\noindent Lensing theory is well developed, and there are many derivations of the relevant equations (e.g. \citeauthor{1999PhDT........23M} \citeyear{1999PhDT........23M}; \citeauthor{2001PhR...340..291B} \citeyear{2001PhR...340..291B}; \citeauthor{2010CQGra..27w3001B} \citeyear{2010CQGra..27w3001B}). \\
Light travels along geodesics, and in the presence of massive bodies, geodesics are curved. As light passes massive bodies, its path is deflected, and a background image will be distorted, magnified, shifted and duplicated. 
Weak gravitational lensing is the term used to describe minimal distortion of the background source with no observable multiplication of the source image. 

For this paper the focus is on linearisable weak lensing, that is, lensing where the distortions are small, and do not vary across the source. This class of lensing can be expressed as a single transformation matrix. It is this feature which forms the basis of the method by which the DSM shear values are determined. We will also restrict ourselves to individual lenses (e.g. galaxies or clusters), however this method is equally valid for measuring the cosmic lensing signal.\\

\noindent If one assumes the length scales of the lensing mass distribution are much smaller than the observer-lens distance, then the thin lens approximation can be used. In the thin lens approximation we project the lensing mass distribution onto a plane perpendicular to the observer's line of sight to obtain the surface mass density:
\begin{equation}
\Sigma(\vec{\xi}) = \int\mathrm{d}\vec{z}\rho(\vec{\xi},\vec{z}),
\end{equation}
where $\vec{\xi} = \sqrt{\vec{x}^2+\vec{y}^2}$ is a set of coordinates in the lens plane, and $\vec{z}$ is the third coordinate parallel to the observer's line of sight. Using the weak field approximation, the deflection angle of light passing the lensing mass is then given by
\begin{equation}
\vec{\alpha}(\vec{\xi}) = \frac{4G}{c^2}\int \mathrm{d}^2\xi'\Sigma(\vec{\xi}')\frac{\vec{\xi}-\vec{\xi}'}{|\vec{\xi}-\vec{\xi}'|^2},
\end{equation}
where $\Sigma(\vec{\xi})$ is the (projected) surface mass density and 
\begin{equation}
M(<\vec{\xi}) = \int \mathrm{d}^2\xi'\Sigma(\vec{\xi}')
\end{equation}
is the mass enclosed within $\vec{\xi}$. The surface mass density can be written in a dimensionless scaled form $\kappa(\xi) = {\Sigma(\xi)}/{\Sigma_{cr}}$, where \begin{equation}
\Sigma_{cr} = \frac{c^{2}}{4\pi G}\frac{D_{s}}{D_{d}D_{ds}} \\
\end{equation}
is the critical surface mass density. $\Sigma\geq\Sigma_{cr}$ yields multiple images (i.e. strong lensing) while $\Sigma\leq\Sigma_{cr}$ gives only one (i.e. flexion and weak lensing).
\\

\noindent The lens equation is
\begin{equation}
\label{eq:lens_eq}
\vec\beta = \vec\theta - \frac{D_\mathrm{ds}}{D_\mathrm{s}}\vec\alpha,
\end{equation}
where $\theta = \xi / D_\mathrm{d}$ is the apparent angular separation of the lens and source, $\beta = \eta/D_\mathrm{s}$ is the true angular separation of the lens and source and $\eta$ is the projected distance between the lens and source in the source plane. $D_\mathrm{ds}$ is the angular diameter distance between the lens and source, $D_\mathrm{s}$ is the angular diameter distance between the observer and source and $D_\mathrm{d}$ is the angular diameter distance between the observer and lens.

\noindent In the thin lens limit, if the bend angle is slowly varying with position, then the lens mapping can be locally linearized. Equation~(\ref{eq:lens_eq}) can then be re-expressed as a coordinate mapping between the lensed and un-lensed coordinate systems\begin{equation}
\vec\beta = \mathcal{A}\,\vec\theta,
\end{equation}
where $\mathcal{A}$ is the Jacobian of transformation and is given by
\begin{equation}
\begin{split}
\label{eq:A_tot}
\mathcal{A}(\vec\theta) & = \frac{\partial\vec\beta}{\partial\vec\theta} = (\delta_{ij} - \frac{\partial^{2}\psi(\vec\theta)}{\partial\theta_{i}\partial\theta_{j}}), \\
 & = \left(\begin{array}{cc}
1-\kappa-\gamma_{1} & -\gamma_{2} \\
-\gamma_{2} & 1-\kappa+\gamma_{1} \\
\end{array}\right), 
\end{split}
\end{equation}
where $\beta$ is the angle to the true source location, $\kappa$ is the convergence (as defined previously), and $\gamma_{1}$ and $\gamma_{2}$ are the shear components given by
\begin{equation}
\gamma_{1} = \frac{1}{2}(\frac{\partial^{2}\psi}{\partial\theta_{1}^{2}}-\frac{\partial^{2}\psi}{\partial\theta_{2}^{2}}),\ \  \gamma_{2}=\frac{\partial^{2}\psi}{\partial\theta_{1}\partial\theta_{2}}, \\
\end{equation}
and $\vec{\gamma} = (\gamma_1,\gamma_2)$. The magnification is given by $\mu=\frac{1}{det\mathcal{A}}=\frac{1}{(1-\kappa)^{2}-|\gamma|^{2}}$. $\mathcal{A}$ is the most general form of the linearised weak lensing transformation, but not the most useful for our purposes. In the weak regime, $\kappa$ has only a small effect as $\Sigma\ll\Sigma_{cr}$. Furthermore, since the DSM algorithm compares the velocity map of a galaxy to reflections of itself, and $\kappa$ produces a spatial magnification only, it will have no effect on the fit. With this in mind, the $\kappa$ term in Equation~(\ref{eq:A_tot}) is discarded for simplicity, so that we can define a reduced transformation
\begin{equation}
\label{eq:A_red}
\mathcal{A_\mathrm{r}} \equiv \left(\begin{array}{cc}
1-\gamma_{1} & -\gamma_{2} \\
-\gamma_{2} & 1+\gamma_{1} \\
\end{array}\right), 
\end{equation} 
and $\gamma_1 = |\vec{\gamma}|\cos(2\theta)$ and $\gamma_2 = |\vec{\gamma}|\sin(2\theta)$. $\mathcal{A}_\mathrm{r}$ can be expressed in a simpler form by selecting a coordinate system such that the shear vector is at a fixed angle with respect to the coordinate axes \citep{2006ApJ...650L..21M}. We choose to define our coordinates such that the shear vector lies at $45^\circ$ to the $y$-axis, to obtain
\begin{equation}
\mathcal{A}_{45} 
= \left(\begin{array}{cc}
1 & -\gamma \\
-\gamma & 1 \\
\end{array}\right), 
\end{equation}
where $\gamma = |\vec{\gamma}|$. It is an important property of $\mathcal{A}_{45}$ that it is invertible and unitary, i.e. 
\begin{equation}
\mathcal{A}_{45}^{-1} = \frac{1}{1-\gamma^2}
\left(\begin{array}{cc}
1 & \gamma \\
\gamma & 1 \\
\end{array}\right), 
\end{equation}
and $\mathcal{A}_{45}^{-1}\mathcal{A}_{45}={A}_{45}\mathcal{A}_{45}^{-1}=I$. To recover the full shear vector in a generalised coordinate system with angle $\theta_t$ to the shear vector, from a measured shear of $\gamma_\mathrm{meas} = |\vec{\gamma}|$, one simply projects onto the new coordinate system:
\begin{equation}
\vec{\gamma} = (\gamma_1,\gamma_2) = (\gamma_\mathrm{meas}\cos(2\theta_t),\gamma_\mathrm{meas}\sin(2\theta_t))\label{eq:total_shear}.
\end{equation}
The angle $\theta_t$ can be obtained by knowing the direction to the lensing mass. In the cases considered in this paper the lensed galaxy is at low redshift and so the direction to the lensing mass can be determined.
\noindent We now have a linearised lensing transformation matrix $\mathcal{A}_{45}$ with which to map our image coordinates according to a shear strength $\gamma$. \\

\noindent To set up the problem, we first define terminology and the relevant coordinates. There are four relevant coordinate systems to consider:
\begin{enumerate}
\item The intrinsic galaxy coordinates in the source plane, prior to lensing: $(m,n)$.
\item The observed galaxy coordinates in the detector plane: $(m^{\prime},n^{\prime})$. The prime notation refers to the observed frame, in which a shear signal is present.
\item The detector frame coordinates: $(x^{\prime},y^{\prime})$. These coordinates correspond to the pixel coordinates of the detector. Note that since they sit in the plane in which a shear signal is present, they are also primed.
\item The intrinsic frame resulting from taking the detector coordinates and projecting them back into the source plane: $(x,y)$.
\end{enumerate}

\noindent While it may at first seem counter-intuitive to use primed notation to denote the observed frame, it is natural to do so considering it is the observed frame that it is lensed, and the intrinsic source plane which is not lensed. Note that at no point are the coordinates of the \itshape{}lensing\upshape{} galaxy referred to. The only coordinate systems used are those of the intrinsic plane of the source galaxy, and the plane of the detector. The detector coordinates are centred on the middle of each pixel, and relate to the un-lensed source plane coordinates $(x,y)$ through 
\begin{equation}
\left[\begin{array}{c}
x^{\prime}  \\
y^{\prime} \\
\end{array}\right]
= \mathcal{A}_{45}
\left[\begin{array}{c}
x  \\
y \\
\end{array}\right]. 
\end{equation}
The detector plane galaxy coordinates are the coordinates of the galaxy as it is observed on the sky, and are related to the intrinsic source plane galaxy coordinates through 
\begin{equation}
\left[\begin{array}{c}
m^{\prime}  \\
n^{\prime} \\
\end{array}\right]
= \mathcal{A}_{45}
\left[\begin{array}{c}
m \\
n \\
\end{array}\right], 
\end{equation}
where the `source plane' refers to the plane in which the true galaxy sits.
$(m,n)$ will be aligned at some angle $\phi$ with respect to the source plane, and will have a different origin to the $(x,y)$ coordinates system, i.e. 
\begin{equation}
\left[\begin{array}{c}
x-x_{0} \\
y-y_{0} \\
\end{array}\right]
= \mathcal{R(\phi)}
\left[\begin{array}{c}
m\\
n\\
\end{array}\right], 
\end{equation}
where $\mathcal{R(\phi)}$ is a rotation matrix, and 
\begin{equation}
\left[\begin{array}{c}
x_{z} \\
y_{z} \\
\end{array}\right]
=
\left[\begin{array}{c}
x-x_{0} \\
y-y_{0} \\
\end{array}\right], 
\end{equation}
where $(x_{0},y_{0})$ are some translation of coordinates, such that in the $(x_{z},y_{z})$ coordinate system the galaxy is centred on the origin.
This relationship between $(x_{z},y_{z})$ and $(m,n)$ will become relevant later when discussing the fitting methods employed by DSM.\\
Finally, since $\mathcal{A}_{45}$ is unitary, to undo a lensing transformation, one simply needs to apply the inverse matrix $\mathcal{A}_{45}^{-1}$ to the lensed coordinates; 
\begin{equation}
\mathcal{A}_{45}^{-1}\left[\begin{array}{c}
x^{\prime} _{z} \\
y^{\prime}_{z} \\
\end{array}\right]
= \mathcal{A}_{45}^{-1} \mathcal{A}_{45}
\left[\begin{array}{c}
x_{z} \\
y_{z} \\
\end{array}\right]
= \left[\begin{array}{c}
x_{z} \\
y_{z} \\
\end{array}\right]. 
\end{equation}

\section{Method}
\label{sec:Methods}
\subsection{The DSM symmetry method:}

\noindent The aim of the DSM algorithm is to use an MCMC fitting algorithm to find the best fit shear value in an observed object, taking as its input a velocity map (V), an error map (E) and a mask (M). This is done by looking for axes of symmetry in the velocity map. As shown in Equation~(\ref{eq:total_shear}), when combined with knowledge of the direction to the lens, the best fit shear value provides a measurement of the total shear field along the line-of-sight to the observed source. A set of such measurements can then be used in a number of scientific applications. The lensing object can be either a cluster or field galaxy.\\ 

\noindent The key idea with this technique is that the projected velocity map of a rotationally supported galaxy is symmetric about two axes (the axis of maximum and of zero projected rotational velocity). We are specifically looking at low redshift background sources, so that we have sufficient angular resolution to apply this algorithm. The low redshift of the source and lens means the direction to the lensing mass is well determined, assuming the foreground luminous galaxy is at the centre of mass of its dark matter halo. Thus we measure the mass of the dark matter halo of the foreground galaxy, through weak shear and possibly flexion.
\\

\noindent For circular orbits (such as in spiral galaxies), the projected velocity onto the plane of the observer is $(v(r)m/r)\sin(\zeta)$, where $r$ is the distance from the centre of the galaxy, $m$ is the projected distance along the major axis and $\zeta$ is the angle between the line-of-sight and the galaxy's rotation axis. This results in the maximum rotational velocity along the $m$ axis, and the minimum along the $n$ axis, with the angle between them being $90^{\circ}$, where $m$ and $n$ are the projected distances along the major and minor axes of the galaxy respectively. A diagram of this geometry is presented in \citet{2006ApJ...650L..21M}, Figure 1.

\noindent This orthogonality implies that we should find three symmetries in the velocity map of a rotating galaxy: symmetrical about the major axis, anti-symmetrical about the minor axis and anti-symmetrical about the diagonal between the two.

\noindent Once a galaxy is lensed however, the shape of the galaxy is distorted without the frequency of the light received at each pixel being changed. Since the rotational velocity of a galaxy is determined via the red- or blue-shifting of a known emission frequency, it follows that the apparent rotational velocity of the galaxy will not change under lensing, but the shape will. In other words the image will be distorted along an axis in the direction of the shear field, so that the velocity axes $m$, $n$ will no longer have an angle of $90^{\circ}$ between them, resulting in the loss of all three symmetries from the image. DSM uses this deviation from symmetry to fit for the shear field that the galaxy light has passed through. For further reading on the topic of velocity maps in rotationally supported galaxies see \citet{1978ARA&A..16..103V}.
\\

\noindent DSM fits for the gravitational lensing shear field which best restores symmetry to the image. Since in the weak lensing regime the light distortion is small, it can be linearised and expressed as a single transformation matrix. Thus if a transformation matrix can be found that restores symmetry to the image, this can be inverted to find the transformation that resulted in the initial distortion, and hence the shear strength.

\noindent To measure the degree of asymmetry in the image, the DSM algorithm uses an MCMC routine with many walkers, which aims to maximise a likelihood function by stepping through the parameter space of possible solutions for $(x^{\prime},y^{\prime})$, $\mathbb{R}$ and $\mathbb{S}$ (the definitions of which are given below).
To do this, the MCMC routine iterates over the following steps: 
\begin{enumerate}
\item Translate the image plane galaxy coordinates $(m^{\prime}, n^{\prime})$, so that the galaxy centre sits at the origin in the image plane coordinate system $(x^{\prime},y^{\prime})$. 
\item Apply the following coordinate transformations to a regular coordinate grid, representing the coordinates of the pixels in a (possibly sheared) velocity map image:
\begin{equation}
\mathbb{S}^{-1}\Rightarrow \mathbb{R}^{-1}\Rightarrow \mathbb{T} \Rightarrow \mathbb{R} \Rightarrow \mathbb{S}, 
\end{equation}
where $\mathbb{S}$ ($\mathbb{S}^{-1}$) is a shear (inverse-shear), $\mathbb{R}$ ($\mathbb{R}^{-1}$) is a rotation (inverse-rotation), and $\mathbb{T}$ is a reflection about one of the three symmetries in the image. The coordinate set resulting from these transformations is used to map the input image to three reflected trial images.
\item In the detector coordinates $(x^{\prime},y^{\prime})$, compute the per-pixel differences between the reflected and reference (i.e. original) images, resulting in three `difference images'.  
\item For each difference image, compute the per-pixel $\chi^{2}$ (giving three sets of values)
\item For each set of per-pixel $\chi^{2}$'s, compute the likelihood value.
\item Add the three sets of per-pixel likelihoods, and then sum over the pixels to obtain the total likelihood for this fit. The shear is obtained from the MCMC maximum likelihood fit. 
\end{enumerate}

\begin{figure*}
\begin{center}
\includegraphics[width=16.5cm]{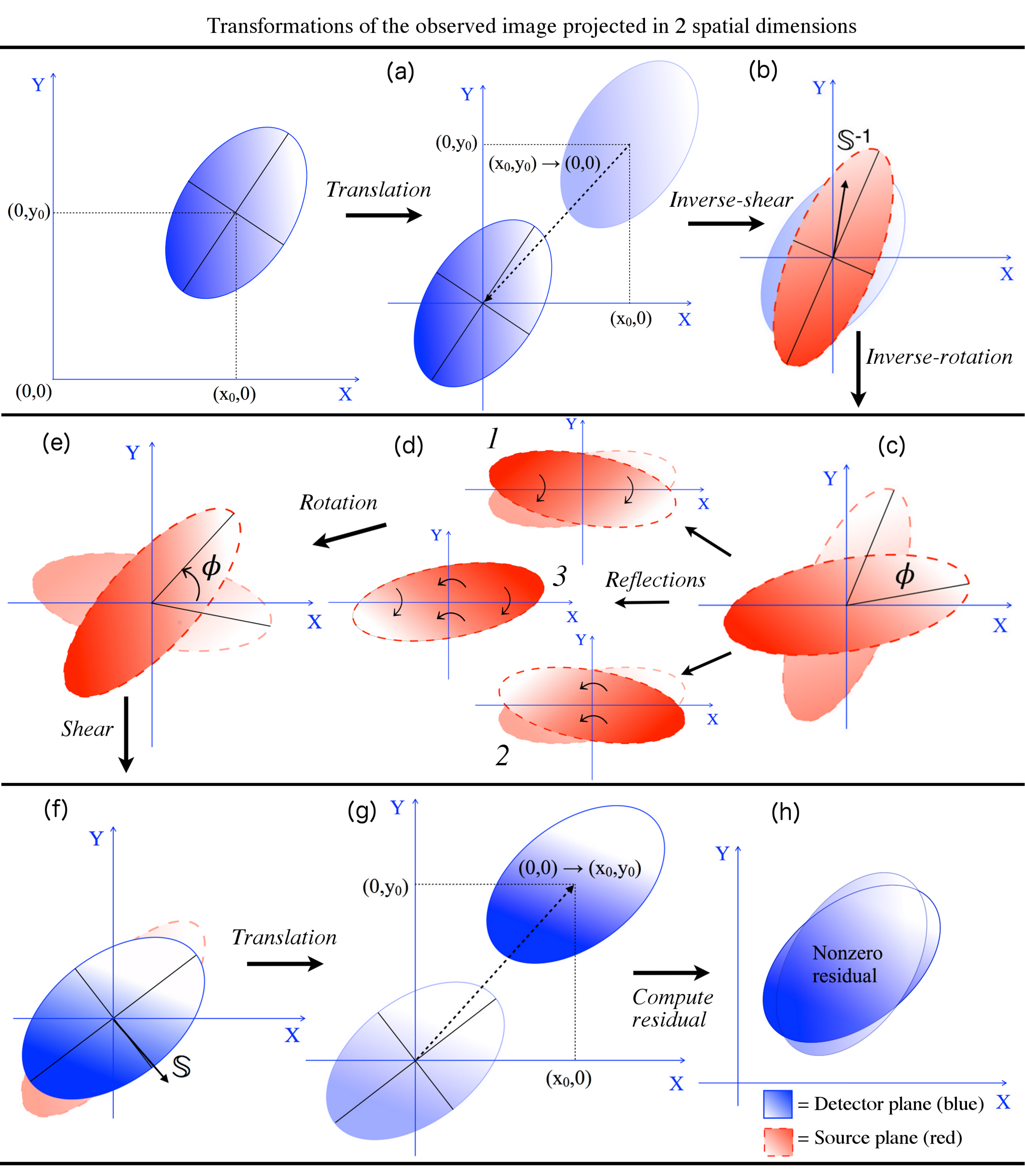}
\caption{A schematic of the transformations in the fitting process. Blue (solid lines) corresponds to the object in the detector plane. Red (dashed lines) indicates the object projected back into the source plane. It is important to understand that while each step is shown individually here, this is for illustrative purposes, and in the DSM algorithm all of the steps shown are done in one transformation. This is in order to minimise interpolation errors. Since the object is initially in the detector plane, and ends up in the same plane, so long as all of the transformations shown here are done in one step (in the order shown), the only point at which interpolation is necessary is in the final step, when computing the residual. The steps are as follows: (a) is a translation to the origin of the detector coordinate system $(x^{\prime},y^{\prime})$ from the detector plane galaxy coordinates $(m^{\prime},n^{\prime})$. (b) is an inverse-shear, taking the object from the detector coordinate system to the un-lensed source plane equivalent, $(x,y)$. (c) is a rotation, the aim of which is to leave the object aligned with the axes of the detector coordinates $(x^{\prime},y^{\prime})$. In (d) reflections are made about the $x$ and $y$ axes of the $(x^{\prime},y^{\prime})$ coordinate system. (e) reverses the rotation applied in (c). (f) re-applies the shear removed in (b), taking the galaxy back to the detector plane. (g) is a translation, taking the galaxy back to its original location, so that it is centred on the $(m^{\prime},n^{\prime})$ coordinate system. (h) is the computation of the residual, and interpolation back onto the detector coordinates $(x^{\prime},y^{\prime})$. The combination of transformations in this series is clearly not the correct set of values, as there is a nonzero residual apparent in (h). }
\label{fig:lensingdiagram}
\end{center}
\end{figure*}

\noindent A diagram of steps (i) to (iii) is shown in Figure~\ref{fig:lensingdiagram}.\\

\noindent 	The structure of the DSM algorithm is now described in detail.

\subsection{The fitting algorithm}

The core of the fitting algorithm, \textsc{trial$\_$fit}, is structured such that each set of trial parameters is input into \textsc{trial$\_$fit}, and a log-likelihood is returned. The set of parameters that results in the maximum likelihood is found by inputting \textsc{trial$\_$fit} into a Monte-Carlo routine.
The module \textsc{emcee} is used in preference over other modules because it employs an ensemble of walkers. This is advantageous for this application since the likelihood surface features local minima (representing incorrect solutions, expected to be a result of discretisation and interpolation errors). The Monte-Carlo routine is invoked by a function \textsc{run$\_$emcee}. The order of processes in \textsc{run$\_$emcee} is as follows:\\

\noindent There are five parameters the algorithms seeks to optimally measure; shear $(\gamma)$, galaxy position angle $(\theta)$, galaxy centre in the $(x,y)$ coordinate system ($x_{0}$ and $y_{0})$, and galaxy central velocity $(v_{0})$. A set of initial guesses for the parameters are specified, followed by the parameters for \textsc{emcee} to run (i.e. number of walkers, number of initial steps to discard (`burn-in steps'), and total number of iterations). A set of initial walker locations is then generated in a `ball' around the initial guess values.\\
A log probability function is defined, which is passed to the fitting parameters. The log probability function first determines whether all the parameters to be fitted lie within a pre-set range of realistic possible values.
\\
\noindent If they do, it calls \textsc{trial$\_$fit}. \textsc{trial$\_$fit} returns the total likelihood and the total $\chi^{2}$.
The burn-in steps are run, and the sampler is then reset so that the burn-in steps aren't counted toward the final fit. \textsc{emcee} is then run for the full number of iterations. \textsc{run$\_$emcee}  then returns the object `sampler' and the array of acceptable walkers. The fitted values of the parameters are obtained by taking the mean of the fitted values of the walkers, then taking the mean of the walkers (in each dimension, for each parameter)

\begin{equation}
\begin{split}
x_{0} &  =  \langle \langle x_{0\,i} \rangle_{j}\rangle, \\
y_{0}  & =  \langle\langle y_{0\,i} \rangle_{j}\rangle, \\
v_{0}  & =  \langle\langle v_{0\,i} \rangle_{j}\rangle, \\
\gamma & =  \langle\langle \gamma_{i} \rangle_{j}\rangle, \\
\theta & =  \langle\langle \theta_{i} \rangle_{j}\rangle \\
\end{split}
\end{equation}

\noindent As mentioned above, the core of the algorithm is \textsc{trial$\_$fit}, which generates the likelihoods, and so some time is taken here to discuss its form. \textsc{trial$\_$fit} takes as its input the set of parameters to test, and the data in the form of a velocity map ($V$), an error map ($E$) and a mask ($M$).\\

\noindent Three reflection matrices are then defined, for reflection of the galaxy coordinates about a major axis, a minor axis perpendicular to the major axis, and about both axes. If the line of reflection is fixed to pass through the origin, and has an angle with respect to the x-axis of $\omega$, then the generalised reflection matrix is \begin{equation}
\begin{split}
\mathbb{T} & = \left(\begin{array}{cc} \cos( 2\omega) & \sin(2\omega) \\
							        \sin(2\omega) & -\cos(2\omega) \end{array}\right). \\
\end{split}				
\end{equation}
\noindent DSM is structured such that $\omega = 0$, so that the reflection axes always align with the detector coordinates $(x_{z}^{\prime},y_{z}^{\prime})$. With $\omega = 0$, the matrices to reflect the coordinates about the major axis, the minor axis and both axes are: 
\begin{itemize}
\item About the major axis: 
\begin{equation}
\begin{split}
\mathbb{T}_{1} & = \left(\begin{array}{cc} 1 & 0 \\
							        0 & -1 \end{array}\right); \\
\end{split}				
\end{equation}

\item About the minor axis:
\begin{equation}
\begin{split}
\mathbb{T}_{2} & = \left(\begin{array}{cc} -1 & 0 \\
							        0 & 1 \end{array}\right);\\	
\end{split}				
\end{equation}

\item About both axes: 
\begin{equation}
\begin{split}
\mathbb{T}_{3} 
& = \mathbb{T}_{1} \mathbb{T}_{2} \\
& = \left(\begin{array}{cc} -1 & 0 \\
				0 & -1 \end{array}\right), \\
\end{split}				
\end{equation}
\end{itemize}
where the subscripts denote the different axes of reflection (1 = major, 2 = minor and 3 = both).\\

\noindent The inverse shear and rotation matrices to be tested are then defined as
\begin{equation}
\mathbb{S}^{-1} = (1 - \gamma_{trial})^{-2}  \left(\begin{array}{cc}  1 & \gamma_{trial} \\
											\gamma_{trial} & 1\end{array}\right), 
\end{equation}
\begin{equation}
\mathbb{R}^{-1} = \left( \begin{array}{cc} \cos ( \phi_{trial} ) & -\sin( \phi_{trial} )\\
						\sin( \phi_{trial} ) & \cos( \phi_{trial} ) \end{array}\right),
\end{equation}
followed by the corresponding shear and rotation matrices

\begin{equation}
\mathbb{S} = \left(\begin{array}{cc} 1 & -\gamma_{trial} \\
							      -\gamma_{trial} & 1\end{array}\right), \\
\end{equation}
\begin{equation}
\begin{split}
\mathbb{R}  & = \left(\begin{array}{cc} \cos(-\phi_{trial}) & -\sin(-\phi_{trial})\\
						           \sin(-\phi_{trial}) & \cos(-\phi_{trial}) \end{array}\right)\\
			& = \left(\begin{array}{cc} \cos(\phi_{trial}) & \sin(\phi_{trial})\\
						           -\sin(\phi_{trial}) & \cos(\phi_{trial}) \end{array}\right),\\
\end{split}
\end{equation}
\noindent where $\gamma_{trial}$ is a trial shear strength and $\phi_{trial}$ is a trial value for the position angle of the object in the source plane.\\

\noindent For each reflection (about the major axis, the minor axis and both), the matrices are multiplied together in the order:
\begin{equation}
\mathbb{F}_{1,2,3} = \mathbb{S\ R\ T}_{1,2,3}\ \mathbb{R}^{-1} \mathbb{S}^{-1}, 
\end{equation}
such that an inverse shear of $\gamma_{trial}$ is the first operation to take place, followed by a rotation by $-\phi_{trial}$, and so on.
This series of matrix transformations is then applied to a set of $x_{z}^{\prime}$ and $y_{z}^{\prime}$ coordinate arrays (i.e. the detector coordinates) to produce trial fit coordinates;
\begin{equation}
\begin{split}
x^{t}_{z\,1,2,3} & = \mathbb{F}_{1,2,3}\, x_{z}^{\prime},\\
y^{t}_{z\,1,2,3} & = \mathbb{F}_{1,2,3}\, y_{z}^{\prime}, 
\end{split}
\end{equation}
which are then translated back to the $(x,y)$ coordinate system 
\begin{equation}
\begin{split}
x^{t}_{1,2,3} & = x^{t}_{z\,1,2,3} + x_{0},\\
y^{t}_{1,2,3} & = y^{t}_{z\,1,2,3} + y_{0}.
\end{split}
\end{equation}
Since these coordinates have been transformed, they will not necessarily be integer values and so it is necessary to interpolate the data when drawing values from the input images with the trial coordinates $(x^{t}_{1,2,3},y^{t}_{1,2,3})$. In the DSM algorithm bilinear interpolation is used. To understand the bilinear interpolation process, it is useful to think of the transformed coordinates as a distorted grid, which we are trying to project back onto a regular grid. In this case each distorted grid square will fall across four regular grid squares, or alternatively each regular grid square will have part of four different distorted pixels fall into it. In order to compute the value of each regular grid square, we need to interpolate between the four pixel values it contains. We can work out what fraction of each distorted grid square $I(x^{t},y^{t})$ falls into our regular grid squares $V(p,q)$, and assign a value to the regular grid square which is equal to 
\begin{equation}
\begin{split}
V(p,q) &= f_{x^{t}}f_{y^{t}}I(x^{t},y^{t})\\ 
&+ f_{x^{t}+1}f_{y^{t}}I(x^{t}+1,y^{t})\\
&+ f_{x^{t}}f_{y^{t}+1}I(x^{t},y^{t}+1)\\ 
&+ f_{x^{t}+1}f_{y^{t}+1}I(x^{t}+1,y^{t}+1),
\end{split}
\end{equation}
where $f_{x^{t}}f_{y^{t}}$ denotes the fraction of the distorted grid square $I(x^{t},y^{t})$ with coordinate $(x^{t},y^{t})$ that falls within the regular grid square $V(p,q)$.
In order to minimise interpolation errors, it is desirable to interpolate as few times as possible. To this end, the trial data is kept in a fractional form until the last step in the fitting process, the computation of the per-pixel $\chi$ values, so that it is only interpolated once. In other words $I(x^{t},y^{t}), I(x^{t}+1,y^{t}), I(x^{t},y^{t}+1)$ and $I(x^{t}+1,y^{t}+1)$ are kept separate and transformed separately. This means that from a single input set of 3 data arrays (velocity map, error map and mask), we have $3\times 3\times 4$ arrays until they are interpolated (3 input images, three reflections and four fractions).  For simplicity therefore, in this paper the four fractional arrays (prior to interpolation) are written as a single array, since they are all treated identically up to the point of interpolation. 

\noindent With this in mind, the transformed (fractional) coordinates are used to create transformed data arrays
\begin{equation}
\begin{split}
V^{t}_{1,2,3} & = V(x^{t}_{1,2,3},y^{t}_{1,2,3}),\\
E^{t}_{1,2,3} & = E(x^{t}_{1,2,3},y^{t}_{1,2,3}),\\
\end{split}
\end{equation}
where $V^{t}$ is the transformed velocity map, $E^{t}$ is the transformed error map, and $M^{t}$ is the transformed mask. 

\noindent A set of three `difference images' are then computed by subtracting the reference velocity map (i.e. original un-lensed image) from each of the transformed maps, along with an additive shift which is a function of $v_{0}$, the form of which is dependent on which axis of symmetry is being used:
\begin{equation}
\Delta V_{1,2,3} = \left\{\begin{array}{c}
V^{t}_{1} - V \\
2 v_{0} - V^{t}_{2,3} - V ,\\
\end{array} \right.
\end{equation}
where the two forms are due to the fact that the sign of the velocity map switches across the y-axis, but stays the same across the x-axis (thus the sign is reversed for reflections about the minor axis and about both axes, but remains the same for reflections about the major axis).\\
 
\noindent The error associated with the difference images is given by
\begin{equation}
\Delta E_{1,2,3} = \sqrt{ (E^{t})^{2}_{1,2,3} + E^{2}}.
\end{equation}
At this point any pixels in the velocity or error maps with bad or missing data are flagged. Any pixels on the edge of the object, where pixels containing velocity data would be interpolated with non-finite values are also flagged. A per-pixel $\chi$ value is then computed for each of the three reflections
\begin{equation}
\chi_{1,2,3} = ( \Delta V_{1,2,3} / \Delta E_{1,2,3} ),\\
\end{equation}
It is at this point that it is most advantageous to interpolate the data, since the next step is to square the data. Interpolating prior to squaring the data reduces the interpolation errors since they will be linear, not squared. The interpolated $\chi$ values are clipped to $\sigma_{max}$, a specified maximum allowed value. Any flagged data points are also penalised and set to $\sigma_{max}$. The setting of a maximum allowed value $\sigma_{max}$ avoids the situation where a few bad pixels with a very large error dominate the fit. Since the shear signal is so small compared to the errors from inaccurate data around the edges of the velocity map, if the $\chi$ values are not clipped the fit would be dominated by minimisation of the residuals in the edge pixels. A check is made by eye to determine the location of the bad (clipped) pixels.
The three interpolated $\chi$ arrays are then added together and squared to obtain the per-pixel $\chi^{2}$ value
\begin{equation}
\chi^{2} = \chi_{1}^{2} + \chi_{2}^{2} + \chi_{3}^{2}.
\end{equation}
The three error arrays $(\Delta E_{1,2,3})$ also need to be interpolated at this point, since they are now squared in order to compute the per-pixel log-likelihood. The likelihood of a given pixel having a given value, assuming gaussian errors, is 
\begin{equation}
L =\sum_i^3{ \frac{1}{\sqrt{2\pi\Delta E_i^2}}\exp(-\frac{\chi_i^2}{2})},
\label{eq:likelihood}
\end{equation}
so the masked per-pixel log-likelihood is 
\begin{equation}
\log L = -\frac{1}{2} [\ \chi^{2} + 2\log( \Delta E_{1}\Delta E_{2}\Delta E_{3}) + 3\log(2\pi)  \ ] \times M.
\end{equation}
Note that the extra error terms arise due to the normalisation in Equation~(\ref{eq:likelihood}, and the log-likelihood has been multiplied by the input mask $M$ to ensure the number of pixels included in the final log-likelihood value is the same for every trial fit. The per-pixel log likelihood, $\log L$, is then summed over the pixels to obtain the total log-likelihood, $\log \mathcal{L}$.
\textsc{trial$\_$fit} then returns $\log \mathcal{L}$, which is used by the log probability function \textsc{logp}.
Clearly the smallest residuals in $\Delta V_{1,2,3}$ will come from trial values which leave the object symmetrical about the three axes of reflection, i.e. the trial values corresponding to the true position angle and shear in the input image. 


\section{Modelling and testing of method}
\label{sec:Modellingandtestingofmethod}

To fully characterise the sensitivity of the method, and the efficacy with which it recovers the true shear in an image, a suite of tests were performed on synthetic data. 
The galaxies were modelled using the following steps: 

\begin{enumerate}
\item Generate a model galaxy, with a lensing signal present;
\item Degrade the velocity and spatial resolution of the image (this is done after the lensing signal is imposed since these properties are chiefly due to seeing and instrumental uncertainties); 
\item Perform an MCMC implementation of the symmetry-search process, including input trial inverse-shears and rotations, and computing the $\chi^{2}$ and likelihood of the trial fit;
\item Repeat the above processes systematically for a range of input parameter values;
\item Assess the robustness of the fitting process as a function of each of the input parameters.  
\end{enumerate}

\subsection{Generating the model data}

The model-generating algorithm, \textsc{make$\_$data}, can be roughly broken into two steps; generating the clean maps, and then adding seeing and velocity measurement error. In order to add the seeing, it is necessary to subsample the data. To achieve this, the data is initially generated to have $M\times N\times n$ pixels, where $n$ is the number of subsamples per larger pixel.\\ 

\noindent The data reduction process for radio data results in a single error value for every pixel in the velocity error map \citep{2001isra.book.....T}. In contrast, optical IFU data is constructed from many individually-measured spectra and so has a different error for every pixel in the velocity map. Since the error maps generated for the model galaxies are constant across the map, they are more analogous to radio data than to optical data. More sophisticated error maps which vary across the galaxy would be required to model optical data. This is left for future publications. \\

\noindent To begin with, empty arrays are created for the velocity map, intensity map, and mask. Arrays of coordinates are then created, running from zero to (x$_{p}$,\ y$_{p}$), with size (x$_{p}\times \sqrt{n}$,\ y$_{p}\times \sqrt{n}$)), where x$_{p}$ and y$_{p}$ are the lengths of the $x$ and $y$ axes of the final images to be made, and $n$ the number of subsamples per pixel to be used when convolving with a Gaussian. These arrays are called $x_{i}$ and $y_{i}$. These coordinates are then inclined about the x-axis
\begin{equation}
y_{incl} = (y_{i} )\sin( \eta ),
\end{equation}
where $\eta$ is the inclination angle. 
These inclined coordinate arrays are then lensed and rotated by the required values
\begin{equation}
\begin{split}
x^{\prime}_{i} &= \mathcal{K} \, x_{i},\\
y^{\prime}_{incl} &= \mathcal{K} \, y_{incl},\\
\end{split}
\end{equation}
where 
\begin{equation} 
\mathcal{K} = \left(\begin{array}{cc}
1 & -\gamma \\
-\gamma & 1 \\
\end{array}\right) 
\left(\begin{array}{cc}
\cos{\theta} & -\sin{\theta} \\
-\sin{\theta} & \cos{\theta} \\
\end{array}\right).\\		
\end{equation}
These coordinates are then used to generate a galaxy image with an exponential surface brightness profile \citep{1970ApJ...160..811F,2008gady.book.....B}
\begin{equation}
I(x,y) = i_0\exp(-2R),
\end{equation}
which gives the following expression for the velocity map 
\begin{equation}
\begin{split}
V(x,y) = v_0& R^2\left[I_0(R)K_0(R)-I_1(R)K_1(R)\right]\\
&\times \cos[\arctan(\frac{x^{\prime}_{i} }{y^{\prime}_{incl} } ) ],
\end{split}
\end{equation}
where $i_0$ and $v_0$ are scaling constants which are set to unity, $K_0$, $K_1$ and $I_0$, $I_1$ are Modified Bessel Functions of the first and second kind respectively, and 
\begin{equation}
R \equiv \frac{\sqrt{{x^{\prime}_{i}}^{2} +  {y^{\prime}_{incl}}^{2}  } }{2 R_{d}},
\end{equation}
where  $R_{d}$ is the disk scale radius.
\\

\noindent This inclination, lensing and rotation procedure assumes the galaxy has no thickness, which is a reasonable approximation since the characteristic scale length of a spiral galaxy disc is typically several orders of magnitude larger than the thickness or scale height \citep{2008PASA...25..184G}. Future implementations of the algorithm will investigate this assumption, and if the effects of thickness are non-negligible it can be incorporated into the modelling. Thickness will be more significant when considering elliptical systems, and if appreciable, will smooth out the shear signal.
\\
A mask is then created with the radius of the mask being set at some multiple of the disk scale radius, $R_{d}$. This mask ensures that only the uniformly-rotating inner parts of the galaxy are used in the fitting algorithm, and the outer regions, which are low signal-to-noise and often disturbed, do not bias the measurement. For real galaxies, the mask size is determined by inspection, including the regular inner regions of the galaxy.  It is noted that there is no real kinematic signature that can mimic the lensing shear signal.\\

\noindent The effects of seeing and imperfect spatial resolution on the velocity and intensity maps were simulated by convolving the flux-weighted velocity map with a Gaussian kernel of width $s$ which when combined with assumptions about model redshift and size, can be expressed in pixels. The Gaussian kernel is given by 
\begin{equation}
G(x,y,s) = \sqrt{ \frac{1}{ 2 \pi s^{2} } }\exp\left[-(\frac{ x^{2} + y^{2}} { 2 s^{2} }  ) \right], 
\end{equation}
so that the flux-weighted degraded image is given by 
\begin{equation}
V_{w}(x,y) = \left[ V(x,y) \times I(x,y) \right] \otimes G(x,y,s).
\end{equation}
The degraded images are then re-binned to the final size (i.e. the array sizes are re-binned from (x$_{p}\times \sqrt{n}$,\ y$_{p}\times \sqrt{n}$) to (x$_{p}$,\ y$_{p}$)), and normalised for the flux-weighting.
\begin{equation}
V_{deg}(x,y) = V_{w}(x,y)/I(x,y)
\end{equation}
Finally, the images have a measurement error (i.e. noise) added to the pixel values. The noise is drawn from a Gaussian distribution peaked around the central velocity, with a width specified as a fraction of the maximum velocity range

\begin{equation}
\begin{split}
V^{\prime}_{deg}(x,y) &= V_{deg}(x,y) + N(V_{0}, V_{frac}),\\
V_{frac} &= \delta V/V_{max},
\end{split}
\end{equation}
where
\begin{equation}
N(V_{0},V_{frac}) =  \frac{1}{ (\sqrt{2 \pi}) V_{frac}  }\exp\left[-\frac{1}{2}(\frac{V_{0}}{V_{frac}})^{2} \right],\\
\end{equation}
where $V^{\prime}_{deg}$ is the 3D image with error in the velocity added, $\delta V$ is the error in the velocity, $V_{frac}$ is the fractional error in the velocity, $V_{0}$ is the central velocity, and $V_{max}$ is the maximum velocity range in the model.

\subsection{Sensitivity tests}

In order to characterise the sensitivity of the DSM algorithm, an ensemble of fits was computed for an exhaustive range of values of the input variables, specified below. The systematic tests were performed on synthetic data generated using \textsc{make$\_$data}, with many realisations of the data fitted, so that while each model had the same measurement error characteristics, the error in each realisation was random. The sensitivity to the following parameters was investigated:

\begin{enumerate}
\item Velocity signal-to-noise ($V_{max}/\delta V$) in the range $V_{max}/\delta V = (2,1000)$. This incorporates the effects on the accuracy of the measured velocity from both spectral resolution and galaxy inclination angle;
\item Sampling resolution (i.e. number of pixels across the galaxy) in the range $p = (2,15)$. This is a proxy for galaxy angular size or redshift;
\item Shear strength in the range of $\gamma = (-0.3, 0.3)$, where this is the same $\gamma$ as defined in Section~\ref{sec:Weaklensing}, to demonstrate the stability of the algorithm for strong shears;
\item Inclination angle $\phi$ in the range $\phi = (10^{\circ},80^{\circ})$. This investigates the effect of a reduced number of pixels and a change in the projected shape of the velocity map with inclination angle. It does not investigate the effect of inclination angle on the accuracy of the velocity; this is taken into account in the investigation of velocity signal-to-noise, discussed above. 
\item Once the other parameters had been investigated, seeing ($s$) in the range $s = (1, 10)$;
\end{enumerate}
Initially, a set of 40 realisations were fit for a set of `fiducial' values of the input parameters, to establish a baseline sensitivity. The values chosen were for a $20$ kpc galaxy, masked beyond $1.5\,$R$_{e}$ at a redshift of $z=0.1$, which gives a mask with a diameter of 15 pixels, and a galaxy with an R$_{e}$ of 5 pixels. The fiducial inclination angle is $\theta_{incl} = 35^\circ$, with a velocity signal-to-noise of $V_{max}/\delta V = 100$. 

\begin{figure}
\begin{center}
\includegraphics[width=8cm]{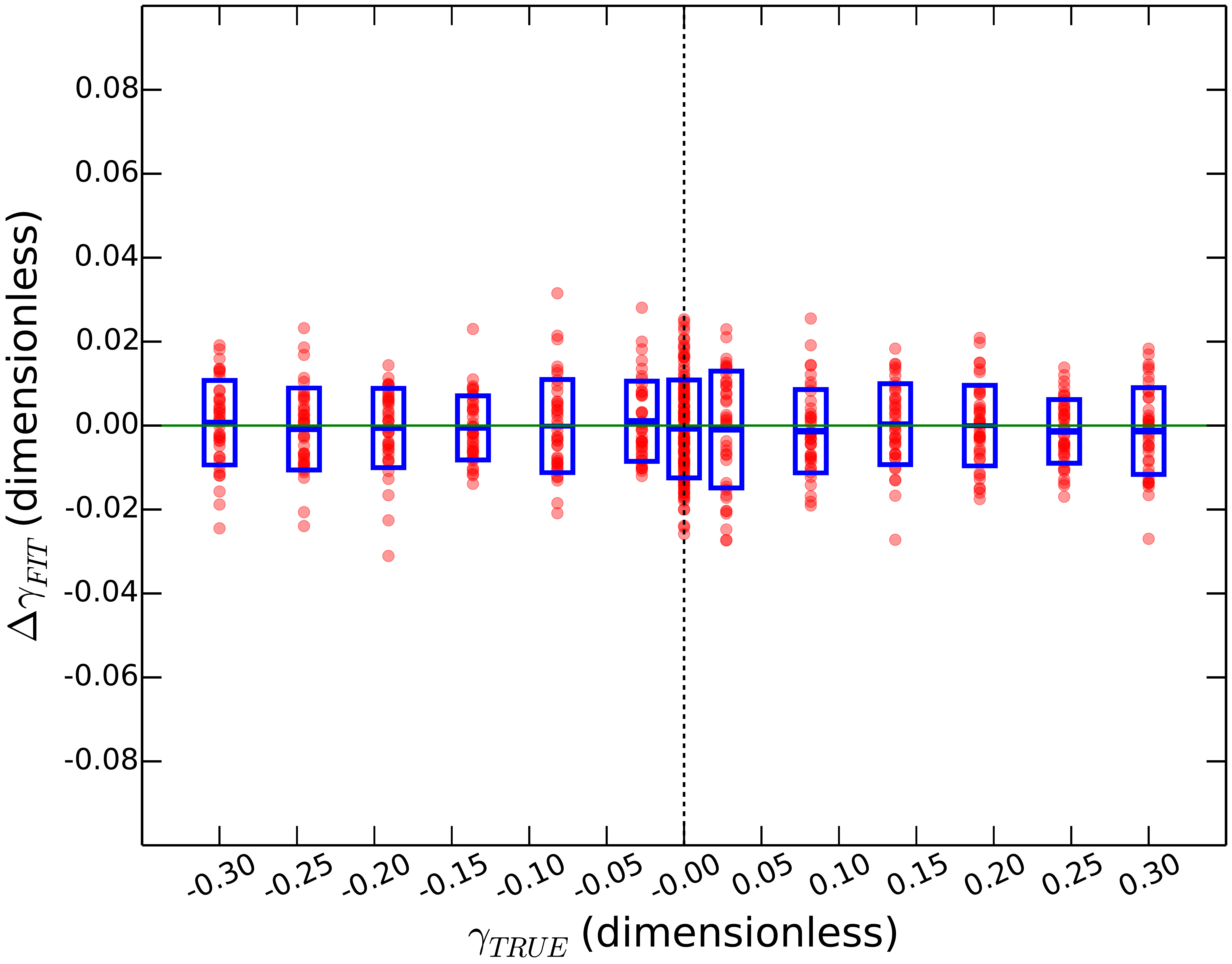}
\caption{The error in the recovered shear values for an esemble of model realisations, as a function of the true input shear.  The parameters used in the galaxy models are $p = 7.5$ pixels, $\theta_{incl} = 35^{\circ}$, $V_{max}/\delta V = 100$. The blue boxes with error bars show the mean and standard deviation in each bin, while the red circles show each individual realisation. There are 40 realisations per bin, with 12 bins over a range $\gamma = (-0.3,0.3)$. }
\label{fig:gammavgamma}
\end{center}
\end{figure}

\begin{figure}
\begin{center}
\includegraphics[width=8cm]{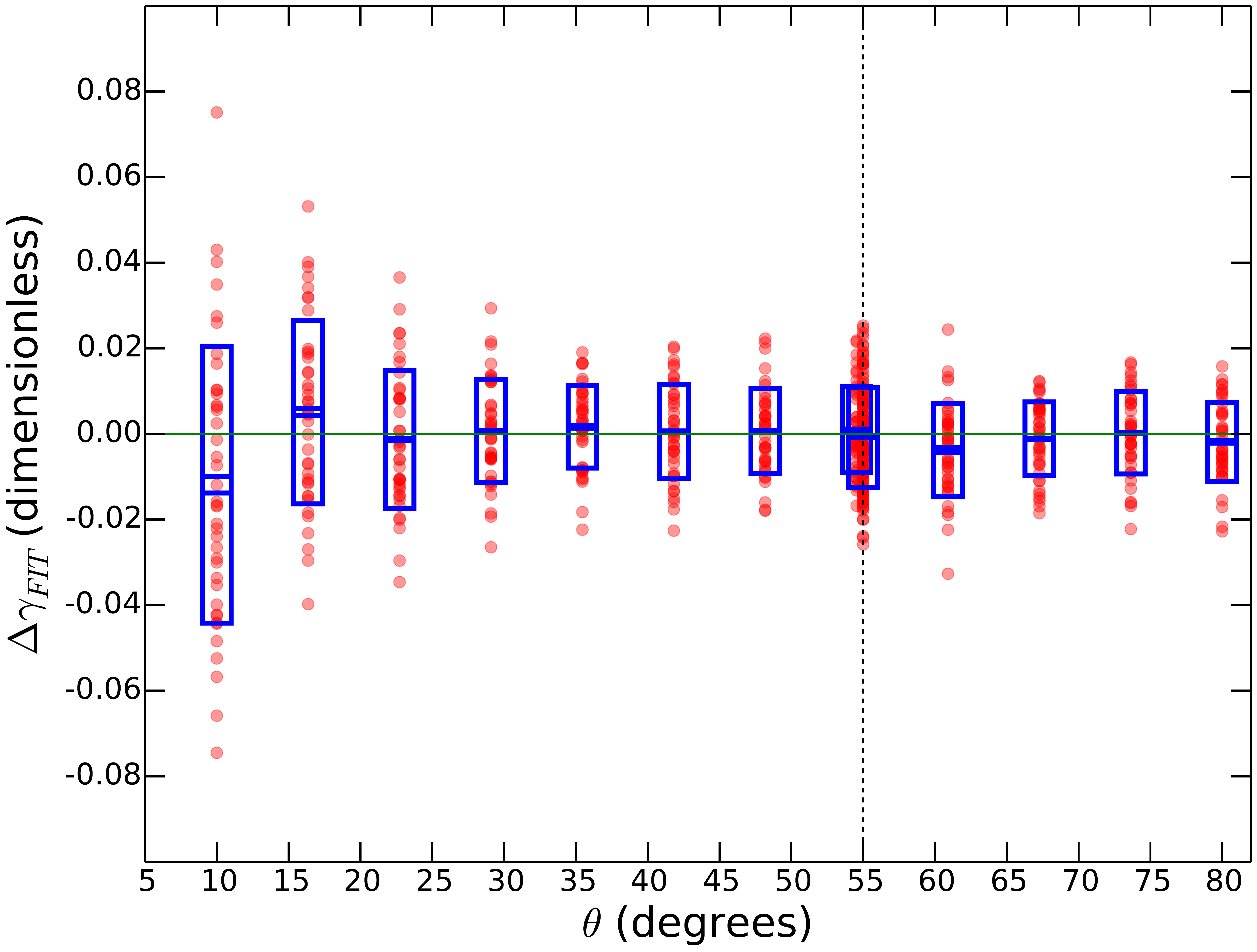}
\caption{The error in the recovered shear values with varying inclination angle, $\theta_{incl}$. The black dashed line shows the fiducial inclination angle value used in the other fits. Symbols and input parameters as in Figure 1.}
\label{fig:gammavinclang}
\end{center}
\end{figure}

\begin{figure}
\begin{center}
\includegraphics[width=8cm]{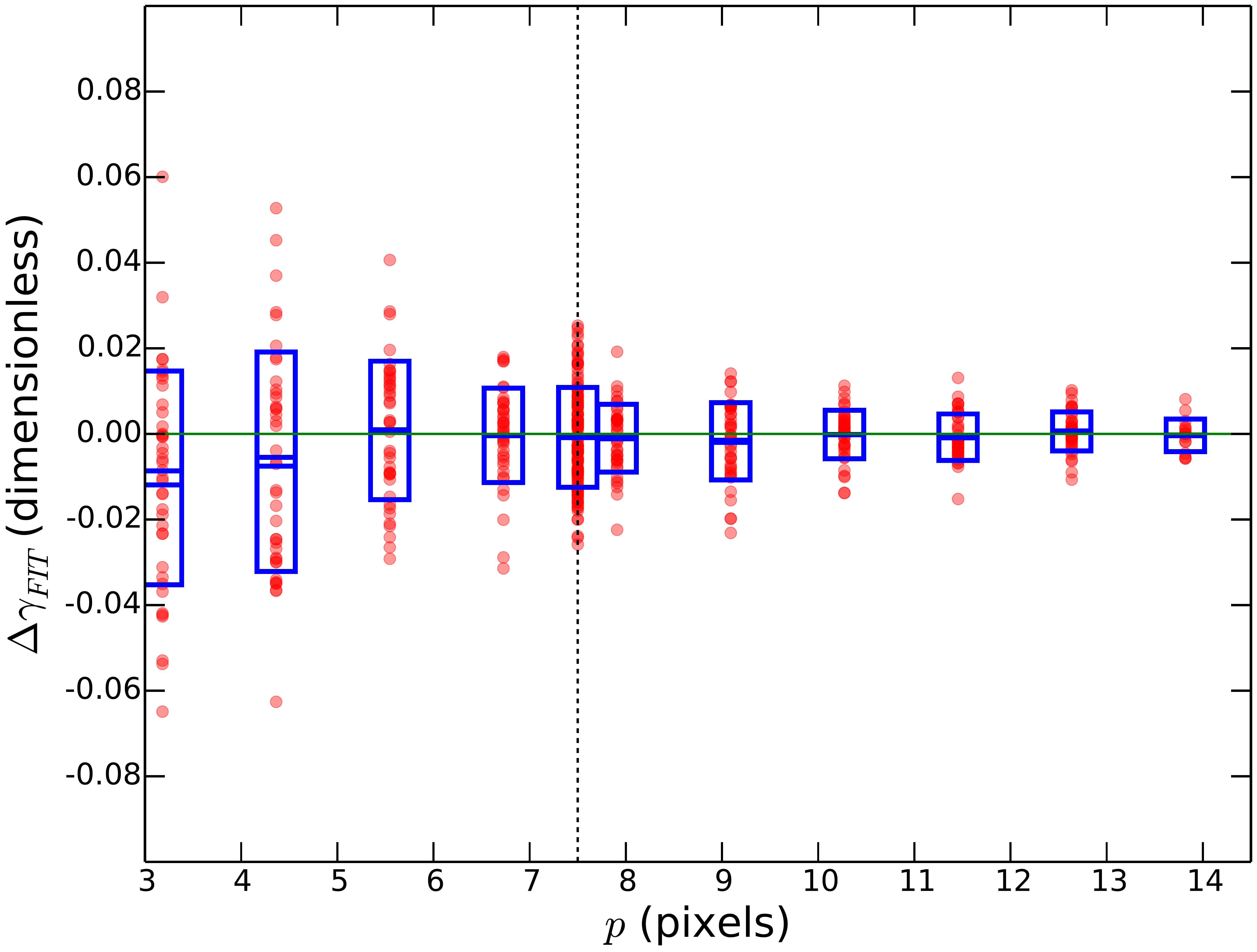}
\caption{The error in the recovered shear values with varying galaxy radius, $p$. The black dashed line shows the fiducial inclination angle value used in the other fits. Symbols and input parameters as in Figure 1. }
\label{fig:gammavgalrad}
\end{center}
\end{figure}

\begin{figure}
\begin{center}
\includegraphics[width=8cm]{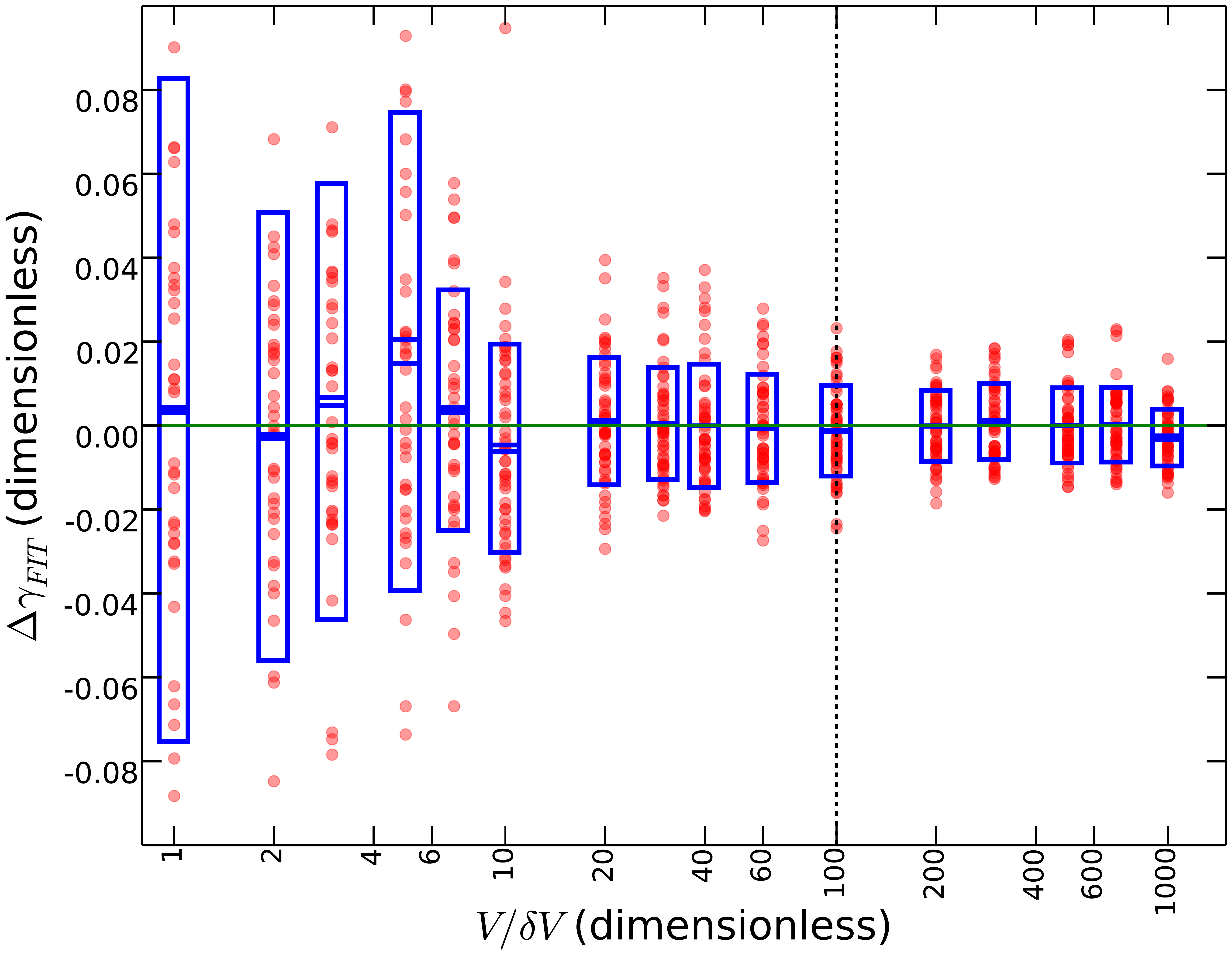}
\caption{The error in the recovered shear values with varying noise, $V_{frac}$. The black dashed line shows the fiducial inclination angle value used in the other fits. Symbols and input parameters as in Figure 1. }
\label{fig:gammavnoise}
\end{center}
\end{figure}

\begin{figure}
\begin{center}
\includegraphics[width=8cm]{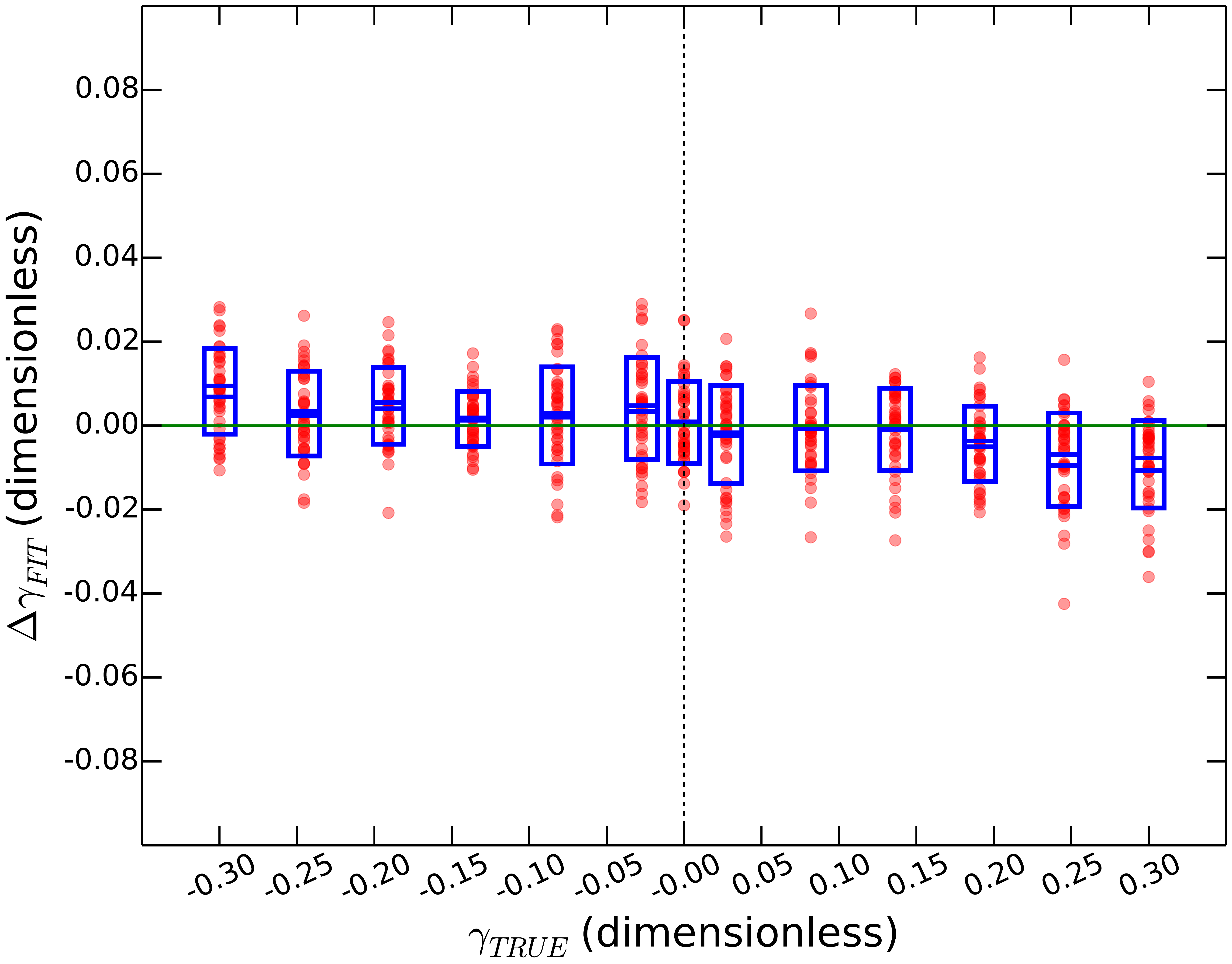}
\caption{The error in the recovered shear values with varying input shear, with seeing of $s=1$, incorporated. Seeing has the effect of decreasing the magnitude of the fitted shear value, compared to the true shear. Symbols and input parameters as in Figure 1. }
\label{fig:gammavgamma_deg=True}
\end{center}
\end{figure}

\begin{figure}
\begin{center}
\includegraphics[width=8cm]{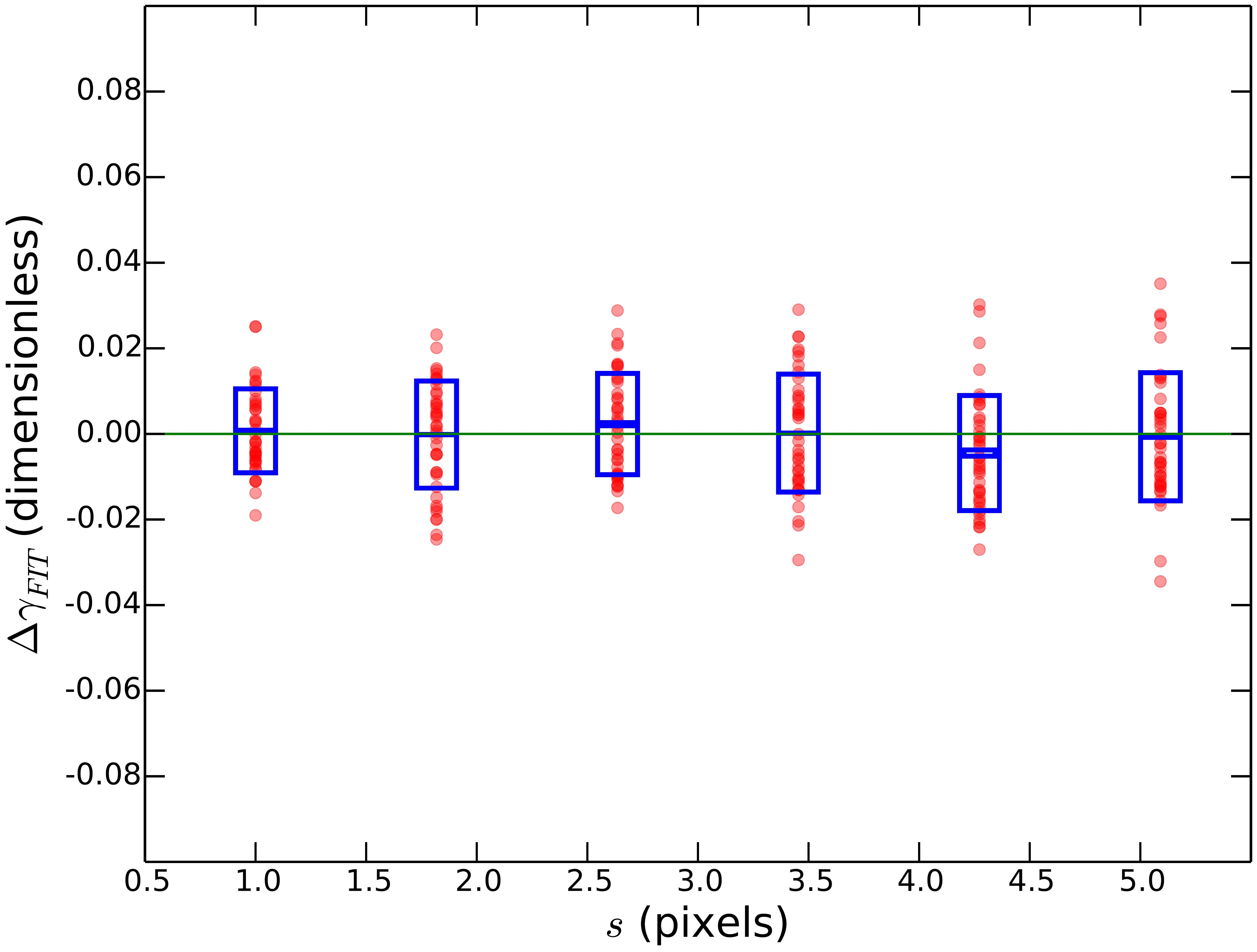}
\caption{The error in the recovered shear values with varying seeing, $s$. Symbols input parameters as in Figure 1. }
\label{fig:FWHMvgamma=0}
\end{center}
\end{figure}

\begin{figure}
\begin{center}
\includegraphics[width=8cm]{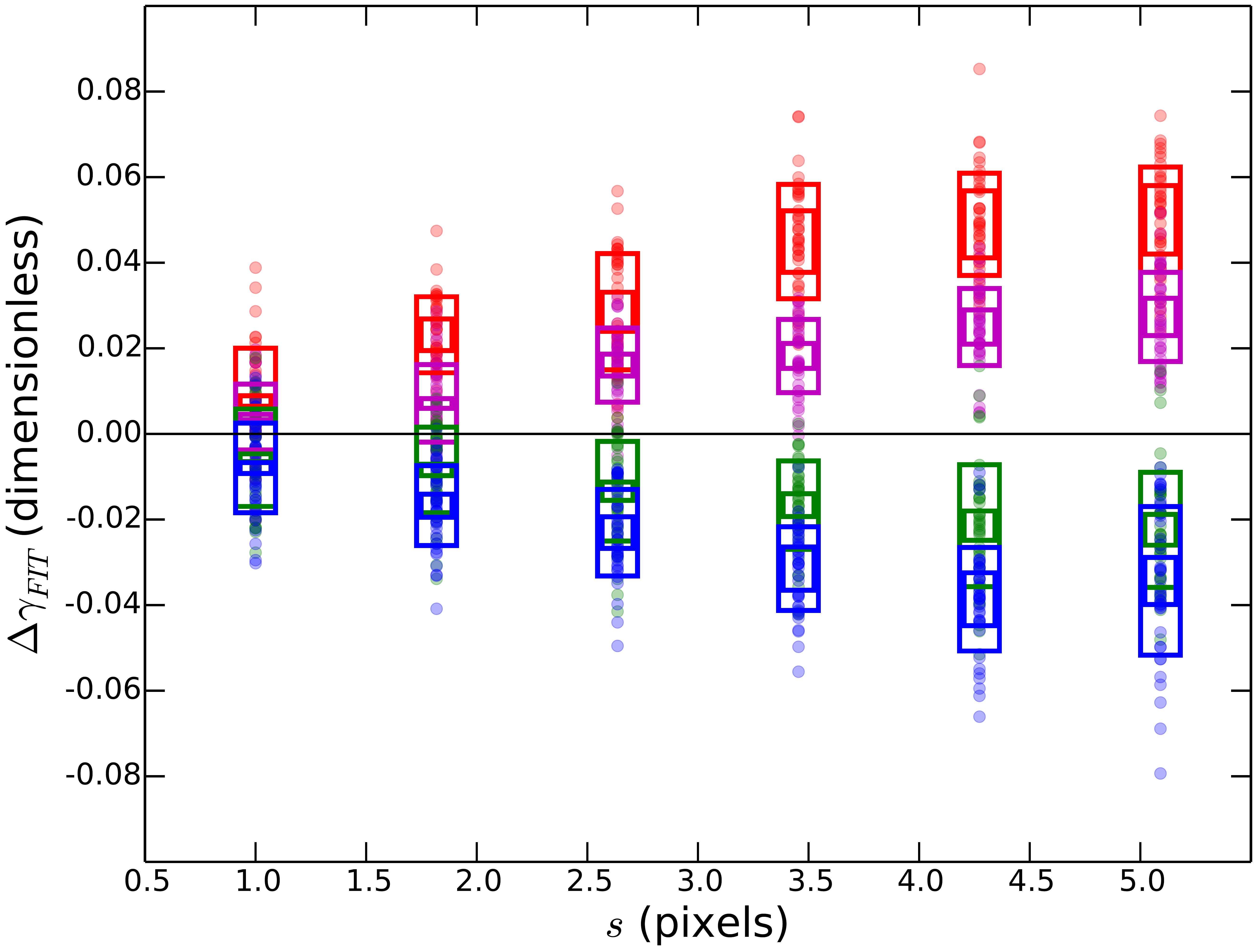}
\caption{The error in the recovered shear values with varying seeing ($s$), for four different input shear values in the model. Red corresponds to an input shear of $\gamma_{in} = 0.3$, magenta corresponds to an input shear of $\gamma_{in} = 0.15$, green corresponds to an input shear of $\gamma_{in} = -0.15$ and blue corresponds to an input shear of $\gamma_{in} = -0.3$. The boxes show the uncertainty of the mean (the inner horizontal lines), and standard deviation (the ends of the boxes), in each bin, while the circles show each individual realisation. For each input value of $\gamma$, there are 40 realisations per bin, with 12 bins over a range $s = (1,10)$. Fiducial values of $p = 7.5$ pixels, $\theta_{incl} = 35^{\circ}$, and $V_{max}/\delta V = 100$ are used, as in the previous models. The slight asymmetry in the plot about the x-axis is a consequence of the position angle (which is held constant in all of the realisations) of the model objects being neither $0^{\circ}$ or $\pm 90^{\circ}$. The error in the fitted shear increases with shear size (and conversely is very small for small shears), so that a typical shear signal ($\gamma\lesssim 0.1$), when measured with good seeing, will have a very small degradation in the fit due to this effect. For example, for a true shear of $0.1$ with a seeing of $s=1$ pixel, the mean fitted shear is $\sim 0.099$, i.e. an error of 0.001.}
\label{fig:gammavFWHM_multi}
\end{center}
\end{figure}

\noindent Once the baseline was established, the behaviour of the fit was investigated as a function of each of the input parameters except seeing. Each parameter was varied over the range mentioned above, while keeping the other parameters fixed at the fiducial values. The results are shown in Figures~\ref{fig:gammavgamma},~\ref{fig:gammavinclang},~\ref{fig:gammavgalrad}, and~\ref{fig:gammavnoise}. For each parameter investigated, 12 values over the range were chosen, and 40 realisations of each value were fitted for. 

\noindent The accuracy of the fit was relatively stable for $\theta_{incl} \gtrsim 30^{\circ}$ (Figure~\ref{fig:gammavinclang}), and is stable for any input shear value in the range $(-0.3 < \gamma < 0.3)$, so long as the shear field is linear.  
The most notable change in sensitivity was with varying velocity signal-to-noise for $V_{max}/\delta V \lesssim 50$ and with the number of pixels across the galaxy for $p \lesssim 8$ pixels. The results of this can be seen in Figures~\ref{fig:gammavgalrad} and~\ref{fig:gammavnoise}. \\ 

\noindent In each figure, the red dots show each individual realisation. The outer blue box shows the standard deviation, while the inner bars in the box show the uncertainty in the mean. Since the number of realisations per bin is reasonably large, the uncertainty on the mean is very small, such that it is hard to see in most bins in most figures. This suggests that we have performed enough realisations for the only uncertainties remaining in this suite of tests to be numerical ones. At worst these uncertainties contribute an error of $\sim$O($1\%$), which is an acceptable value. The exception to the above observation is the systematic bias introduced by seeing, which is discussed below. \\

\noindent Once the effects of noise, shear strength, inclination angle and sampling resolution had been investigated, seeing was introduced, and its effect on the recovered shear accuracy was investigated by varying the input shear while holding the seeing fixed at $s = 1$ (corresponding to a spatial degradation kernel FWHM of $1$ pixel), and then by varying the seeing while holding the input shear fixed at $\gamma = 0$. The results are shown in Figures~\ref{fig:gammavgamma_deg=True} and \ref{fig:FWHMvgamma=0} respectively. 

\begin{table*}
\caption{ Guideline values for the investigated parameters, within which the DSM algorithm performs accurately. The ``Range investigated'' column shows the range in which each parameter was investigated. The ``Lower limit'' and ``Upper limit'' columns give the limits in which a realistic fit could be obtained, given the range investigated. The ``Best fit'' column gives the value of each parameter which resulted in the most accurate fit in the range investigated.
}
\centering 
\begin{tabular}{c c c c c }
\hline
Parameter & Range investigated & Lower limit & Upper limit & Best fit \\ [0.5ex]
$\gamma$ (unitless) & $ -0.3 < \gamma < 0.3 $ & -0.3 & 0.3 & 0.0  \\
$\theta_{incl}$ (degrees) & $ 10 <  \theta_{incl} < 80 $ & 30 & 80 & 45 \\ 
$V_{max}/\delta V $ (unitless) & $ 1 < V_{max}/\delta V < 1000 $ & 50 & none & $1000$ \\
$p$ (pixels)  & $ 2 < p <15 $ & 7.5 & none & $15$ \\
$s$ (pixels)  & $ 1 < s < 10 $  & none & $4$ & $1$ \\[1ex]
\hline
\end{tabular}
\label{table:fitranges}
\end{table*}

\noindent The accuracy of the fit varies little with the size of the degradation kernel (Figure~\ref{fig:FWHMvgamma=0}) for a fixed input shear of $\gamma = 0$, however if the input shear is varied with fixed seeing, it becomes apparent that seeing introduces a bias into the shear signal for any nonzero shear (Figure~\ref{fig:gammavgamma_deg=True}). This is because the seeing `smears' the velocity map, washing out the shear signal, and resulting in the measured shear having a smaller magnitude. \citet{1995ApJ...449..460K} have developed techniques to reduce the impact of such smearing in the case of 2D images in traditional weak lensing applications by attempting to deconvolve the PSF from the image. It may be possible to implement similar techniques for velocity maps and 3D data in general, however we leave an investigation of this possibility to future work.

\noindent To further investigate the effect of seeing on nonzero shears, a second set of model realisations were fit, over a range of degradation kernel sizes for four non-zero shear values; $\gamma = -0.3, -0.15, 0.15$, and $0.3$. As can be seen in Figure~\ref{fig:gammavFWHM_multi}, the effect of seeing is to reduce the magnitude of the shear present in the data, and so increase the error. For a true shear which is positive (negative), increasing seeing give an increasingly negative (positive) error. This effect seems to saturate for very high values of $s$, corresponding to very poor seeing, and is relatively small for all realistic choices of seeing. For a seeing corresponding to a degradation kernel of $s = 2.5$, the error in the fit due to seeing is $\sim$O($2.5\%$) at worst (i.e. in the case of the strongest possible shear). The error in the fitted shear increases with shear size (and conversely is very small for small shears), so that a typical shear signal ($\gamma\lesssim0.1$), when measured with good seeing, will have a very small degradation in the fit due to this effect ($\lesssim 0.1\%$). In light of this, this effect was not seen as being of immediate concern, so long as the seeing conditions for observing are reasonable. The slight positive bias in Figure~\ref{fig:gammavFWHM_multi} is due to the non-alignment of the projected distance between the lens and source and the position angle of the source, and is an expected effect. The uncertainty of the mean in this set of fits is larger than in the previous suite of tests because there are fewer realisations per bin in this case.\\

\noindent Simulations were also undertaken to determine whether ellipticity in the point spread function (PSF) might  bias the measurement of the shear using the DSM algorithm. A mock dataset was created, in which each object was convolved with a PSF with an ellipticity of $e = 0.5$. This is an extreme choice of PSF ellipticity, far greater than would be expected for any current telescopes. In spite of this however, the introduction of ellipticity to the PSF made a negligible ($< 1\%$) difference to the accuracy with which the DSM algorithm recovered the shears present in the mock dataset objects. Furthermore, no dependence between the PSF ellipse orientation and galaxy position angle was observed.\\

\noindent Based on the errors per bin in the set of simulations described above, guideline acceptable fit ranges were specified for each of the input parameters. These define the parameter space in which the DSM method as currently implemented will be useful and effective.  The guideline values for each parameter are presented in Table~\ref{table:fitranges}.

\section{Application to real data}
\label{sec:Applicationtorealdata}
The requirements for the application of the DSM algorithm to a background galaxy are the following: 
\begin{itemize}
\item A reasonable expectation that weak lensing may be applicable, vis an angular distance between the foreground and background galaxies that suggests an estimated shear $\gtrsim 0.01$.
\item The background galaxy has a suitable projection to the line-of-sight. Measurement of the shear from the velocity map is not possible at inclination angles at or close to $\theta_\mathrm{incl} = 0^\circ$ and $\theta_\mathrm{incl} = 90^\circ$. The measurement accuracy of the velocity map is minimum at these angles, and maximum at $\theta \simeq 35^\circ$. This effect is taken into account in our investigation of velocity signal-to-noise in the previous section. 
\item Ideally, the background galaxy is relatively isolated so that the disk motions are circular and undisturbed.
\item The projected separation between the lens (foreground) and source (background) galaxies does not coincide with either the major or minor axes of the source galaxy.
\item A measurable and regular rotation.  In practice, this means that at high redshift there will be fewer suitable galaxies to which the DSM method can be applied.
\end{itemize}
Non-circular motions in the disk of the background galaxy, such as tidal distortions or kinematic motions due to a bar, will result in well-defined signatures in the DSM fitting metric, and cannot be confused with the shear signal.  In such cases it is not possible to fit for a small shear.

\noindent In order to have confidence that DSM is fitting for true shear signals in lensed data, one needs to be reassured that where there is no lensing signal present, DSM recovers a null result. In the previous section DSM's ability to accurately recover a null result in model data was demonstrated, however there are characteristics of real objects that were not included in the models. Examples of this include warp instabilities in the disks of galaxies, barred galaxies, `clumpiness' and non-uniform structure within galactic disks, and turbulence in the outer regions of the galaxy (of particular relevance for radio data).

\noindent To assess how well a null result is recovered in un-lensed observational data, fits were attempted for a shear signal in the velocity maps of very low redshift galaxies. Since the probability of an object being lensed increases with distance (as the projected density of intervening matter increases), then the chance of a very low redshift galaxy being lensed is vanishingly small.  With this in mind the shear was fit in a number of galaxies in the nearby universe, using data at radio wavelengths. Radio data was chosen for this first test since the models investigated in section ~\ref{sec:Modellingandtestingofmethod} have properties analogous to that found in HI data cubes (specifically that the error maps are constant, rather than varying across the image). An investigation of optical integral field spectrograph data is left for future work. 

\noindent Both `regular' and `complex' objects were fitted to. `Regular' objects are those that have a smooth velocity map with no warps, bars or clumps present. `Complex' objects are those with warps apparent, barred galaxies, and those with clumpy structure and generally complex velocity profiles. \\

\begin{figure}
\begin{center}
\includegraphics[width=8.5cm]{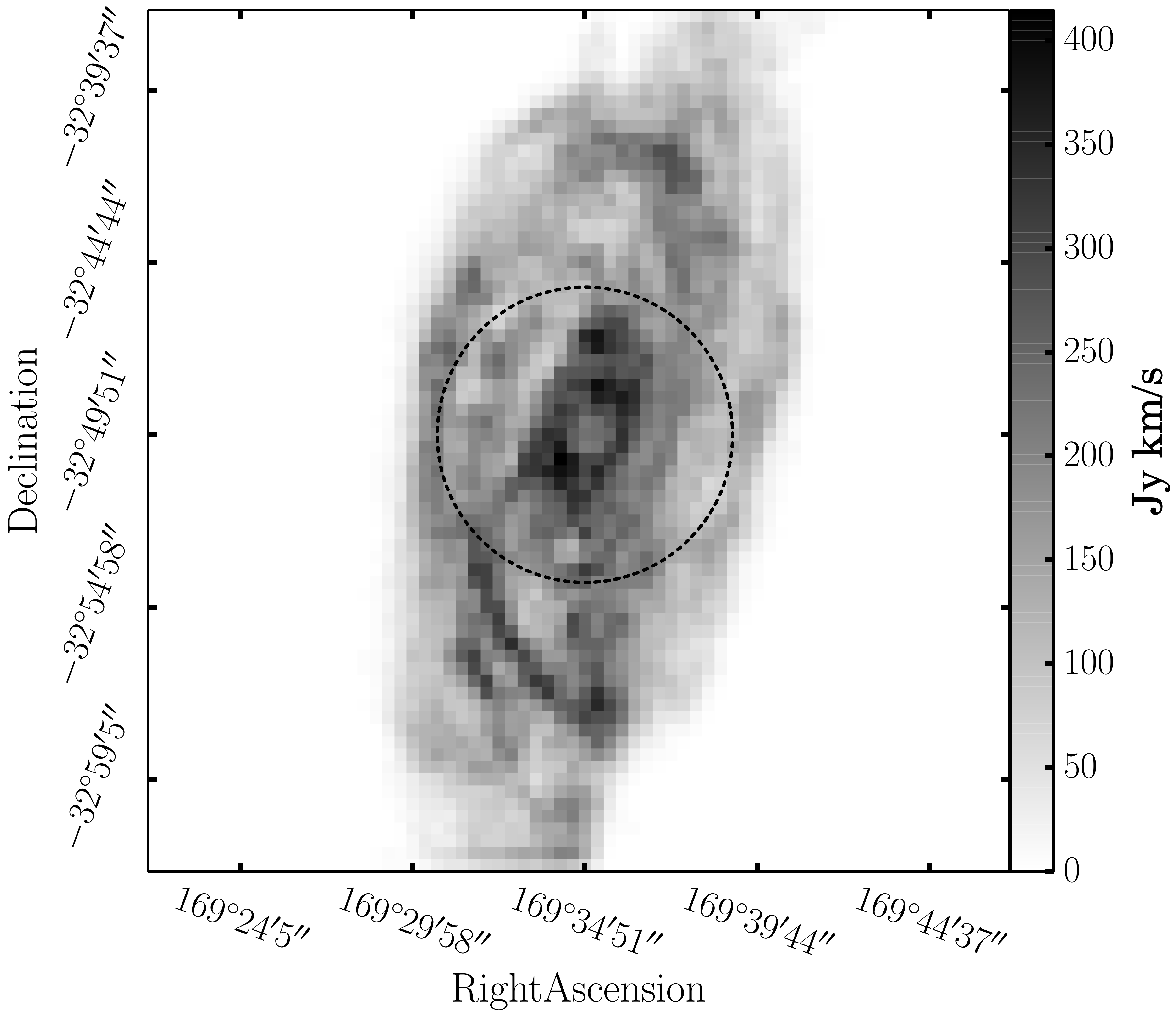}
\caption{ The zeroth moment (i.e the HI emission image) of NGC 3621 from THINGS. The mask radius is shown by the black line. The Fitted parametes for this galaxy are given in Tables~\ref{table:realfits} and~\ref{table:unmaskedfits} for the masked and unmasked fits respectively.}
\label{fig:NGC3621_0}
\end{center}
\end{figure}

\begin{figure}
\begin{center}
\includegraphics[width=8.5cm]{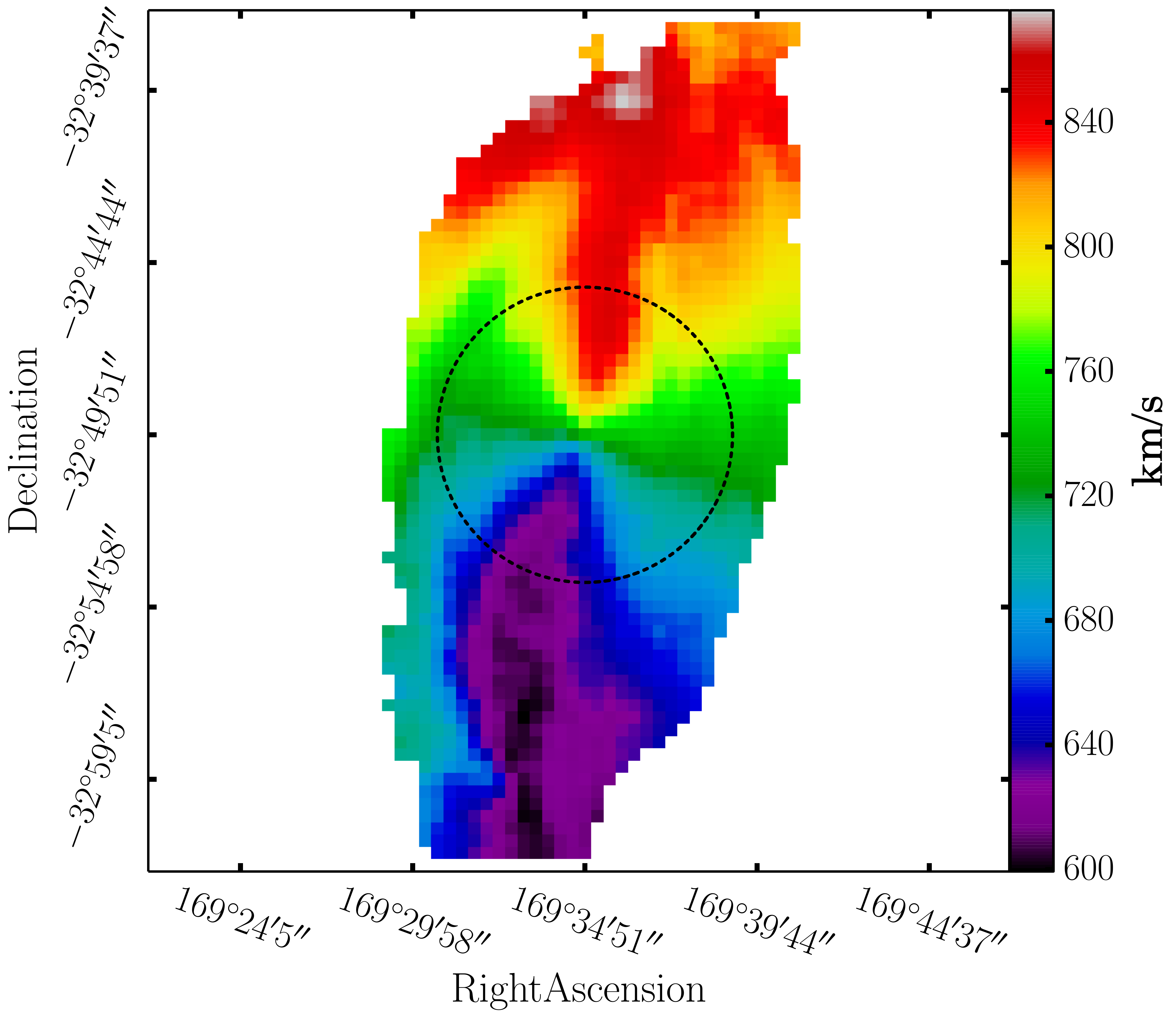}
\caption{ The first moment (i.e. the velocity map) of NGC 3621 from THINGS, with the mask radius shown by the black line. The rotational velocities are based on the HI emission line. The Fitted parametes for this galaxy are given in Tables~\ref{table:realfits} and~\ref{table:unmaskedfits} for the masked and unmasked fits respectively. }
\label{fig:NGC3621_1}
\end{center}
\end{figure}

\begin{figure}
\begin{center}
\includegraphics[width=8.5cm]{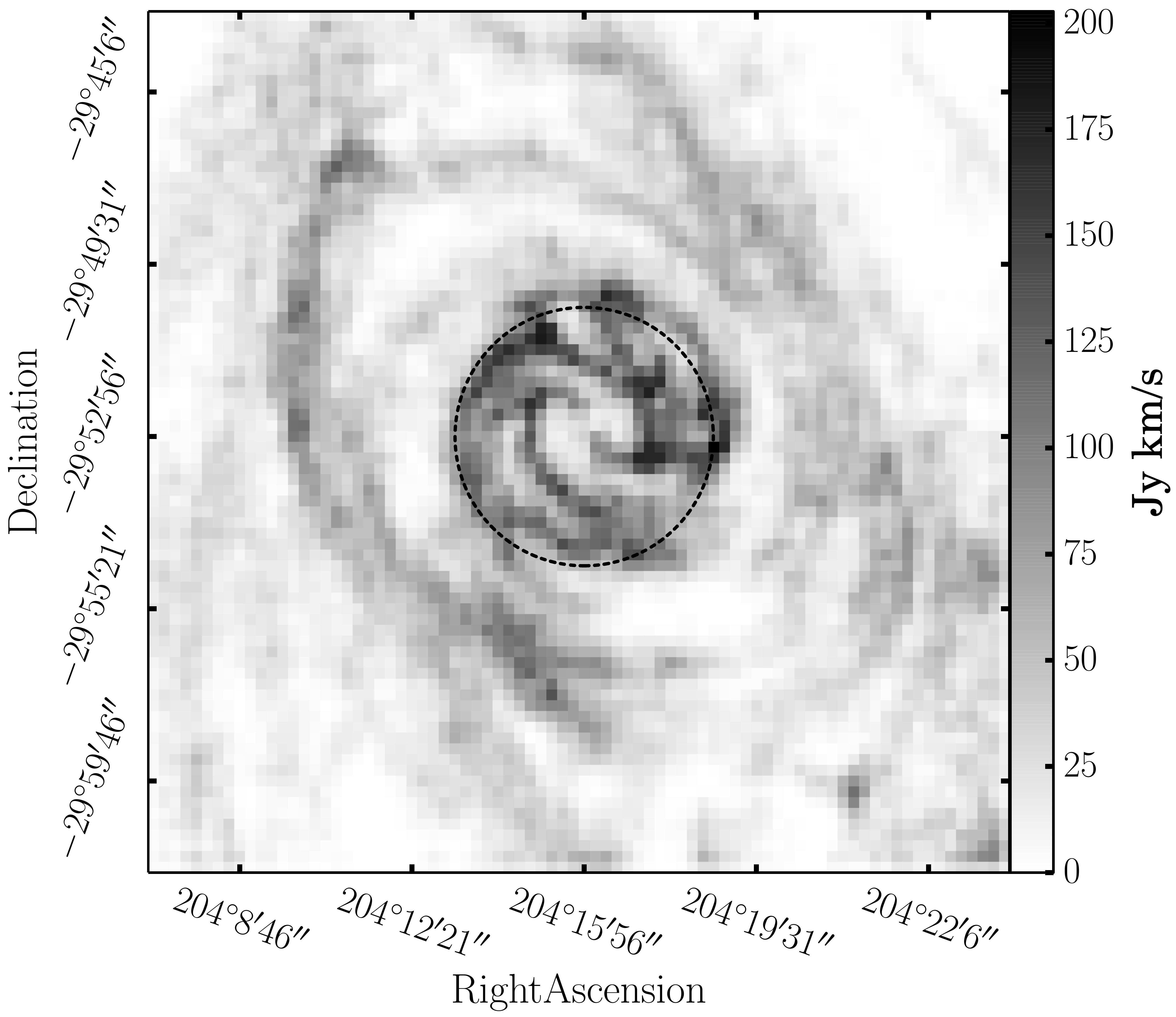}
\caption{The zeroth moment (i.e the HI emission image) of NGC 5236 from THINGS. The mask radius is shown by the black line. The Fitted parametes for this galaxy are given in Tables~\ref{table:realfits} and~\ref{table:unmaskedfits} for the masked and unmasked fits respectively. }
\label{fig:NGC5236_0}
\end{center}
\end{figure}

\begin{figure}
\begin{center}
\includegraphics[width=8.5cm]{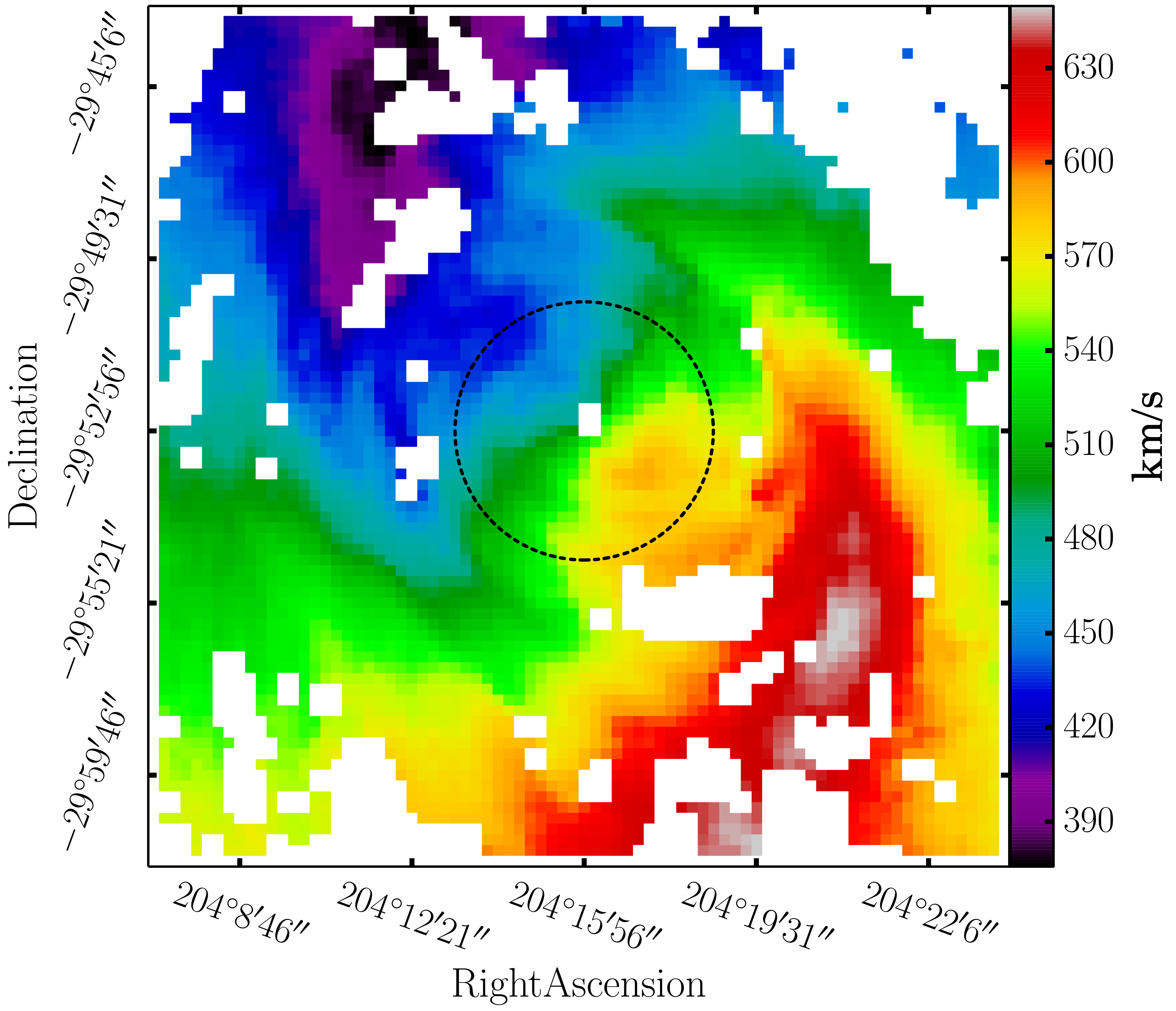}
\caption{The first moment (i.e. the velocity map) of NGC 5236 from THINGS, with the mask radius shown by the black line. The Fitted parametes for this galaxy are given in Tables~\ref{table:realfits} and~\ref{table:unmaskedfits} for the masked and unmasked fits respectively. }
\label{fig:NGC5236_1}
\end{center}
\end{figure}

\noindent Data from The HI Nearby Galaxy Survey (THINGS) \citep{2008AJ....136.2563W} was used, which provides very high angular and velocity resolution ($7^{\prime\prime}$ and $5\,$km$\,$s$^{-1}$ respectively) observations of 34 objects with distances in the range $3<\,$D$\,<15\,$Mpc ($0.0007<z<0.00355$).
Since the velocity maps from THINGS are obtained from measurements in HI, they trace the HI abundance in the galaxies. The HI component of most galaxies extends out to much larger radii than the optical component, into regions where there is more turbulence and the disk is more likely to be warped. For this reason it is advantageous to mask this data beyond the extent of the optical component. Two objects were fitted from THINGS; NGC 3621 with morphological type SA(s)d, and NGC 5236 (M83) with morphological type SAB(s)c. The morphology of these two choices was of particular relevance, as it allowed for the effect of bars on the accuracy of the fit to be investigated. 
The HI emission and velocity map of NGC 3621 are shown in Figures~\ref{fig:NGC3621_0} and~\ref{fig:NGC3621_1}, and those of NGC5236 are shown in Figures~\ref{fig:NGC5236_0} and~\ref{fig:NGC5236_1}. In both cases the mask radius is 12 pixels and is shown by the black line.\\

\noindent Two velocity maps were fitted from THINGS; one barred (NGC 5236) and one not barred (NGC 3621). Each were fitted with the same input parameters for the Monte-Carlo routine; $60$ walkers, $200$ burn in steps, followed by $300$ fitting steps. The fits for each object are presented in Table~\ref{table:realfits}.\\

\noindent Initial guesses for the fit were obtained as follows: For the coordinates of the galaxy centre, the mean in the $x$ and $y$ positions of all nonzero pixels was used. For the central velocity, the mean of all nonzero pixels was taken. The position angle was estimated by eye. For all the fits an initial shear guess of $\gamma = 0$ was used. So long as it was given sufficient time to converge, the algorithm was reasonably robust to poor initial guesses. With poor guesses up to 600 burn in steps were required to obtain a fit of equivalent accuracy. Circular masks were added at approximately the extent of the regular velocity field, in the inner regions of the objects.  This is roughly $1/3$ of the radio extent for both NGC 3621 and NGC 5236. The use of circular masks, necessitated to eliminate bias from complex extended structure, discourages fits involving strong distortions, and therefore damps the fitted shear values. This is an effect that needs to be addressed, however it is beyond the scope of the current analysis.\\

\begin{table*}
\caption{Fitted parameters for the (masked) THINGS. Fits were obtained with a Monte-Carlo paramter space search using $60$ walkers, $200$ burn-in steps and $300$ iterations.}
\centering 
\begin{tabular}{c c c c c c c c c c c}
\hline
 & $\gamma$ & $\delta \gamma$ & $\phi$  & $\delta \phi$  & $x_{0}$ & $\delta x_{0}$ & $y_{0}$ & $\delta y_{0}$ & $V_{0}$ & $\delta V_{0}$ \\ 
\textbf{Object} & (unitless) & (unitless) & (degrees) & (degrees) &  (pixels) &  (pixels) &  (pixels) &  (pixels) &  (km$\,$s$^{-1}$) &  (km$\,$s$^{-1}$)\\[0.5ex]
\hline
NGC 3621 & $-7.91\times 10^{-5}$ & $7.75\times 10^{-4}$ & -9.65 & $7.28\times 10^{-2}$ & 35.00 & $9.1\times 10^{-3}$ & 36.00 & $8.7\times 10^{-3}$ & 727.78 & 1.262 \\
NGC 5236 & $5.22\times 10^{-5}$ & $1.85\times 10^{-3}$ & -119.70 & $1.46\times 10^{-1}$ & 41.03 & $1.2\times 10^{-2}$ & 40.97 & $1.2\times 10^{-2}$ & 509.56 & 1.298 \\  
[0.9ex]
\hline
\end{tabular}
\label{table:realfits}
\end{table*}

\noindent As can be seen the accuracy with which the fits have recovered the expected null result is most affected by the application of a mask to the object. The warps and turbulence prevalent in the outer disks of even un-barred THINGS objects result in biasing of the fits, however masking to the optical radius of the object rectifies this. To illustrate the degree of improvement provided by this masking process, the THINGS objects were fitted for again without masking. The results of this are presented in Table~\ref{table:unmaskedfits}. It is clear from this table that masking to remove spurious signal from complex velocity structures, and retaining only the inner regular region of the velocity map is critical for obtaining good fits. 

\begin{table}
\caption{ Fitted parameters for the un-masked THINGS galaxies. Fits were obtained in the same manner as those presented in Table~\ref{table:realfits}, except that the objects were not masked.}
\centering 
\begin{tabular}{c c c }
\hline
Parameter & NGC 3621 & NGC 5236 \\ 
[0.5ex]
\hline
$\gamma \pm \delta \gamma$ (unitless) & $0.049 \pm 0.0057$ & $-0.234 \pm 0.2587$ \\
$\phi\pm\delta \phi$ (degrees) & $-9.66\pm 0.415$ & $-135.11\pm 7.414$ \\  
$x_{0}\pm\delta x_{0}$ (pixels) & $34.83\pm 0.069$ & $53.13\pm 13.217$ \\  
$y_{0}\pm\delta y_{0}$ (pixels) & $35.31\pm 0.021$ & $23.86\pm 13.513$ \\  
$V_{0}\pm\delta V_{0}$ (km$\,$s$^{-1}$) & $730.96\pm 0.580$ & $377.14\pm 91.497$ \\ 
\hline
\end{tabular}
\label{table:unmaskedfits}
\end{table}

\section{Conclusions}
\label{sec:Conclusions}

A new method for measuring gravitational shear from weak lensing has been described and tested. 
The method is based on the premise that symmetries in the velocity map are lost when it is lensed.  DSM describes a process by which the deviation from symmetry might be measured and a linearised shear field determined. Systematic tests were performed to investigate the robustness of the algorithm, and requirements on the data quality for a high fidelity measurement. These tests also demonstrated the ability of the algorithm to recover a null result in synthetic data. The algorithm was then tested on images of nearby galaxies from the THINGS radio survey. A null result was obtained after appropriate masking of the image. The results of these investigations, presented in Tables~\ref{table:fitranges} and~\ref{table:realfits}, have demonstrated that the DSM algorithm will recover a null result, both in simulated and real data.  The range of parameters within which the DSM algorithm will be useful has also been established. These results are strong motivation to develop the DSM algorithm further. The first step has been made towards quantifying the errors and uncertainties expected when DSM is used to recover a shear signal from observational data. Future work will investigate velocity maps from higher redshift galaxies that have a high probability of being lensed to determine whether a non-zero shear signal is observed. This technique is most effective where the background source is an isolated, undisturbed, stably-rotating disk. We have found that such systems not only exist but are reasonably easy to find (de Burgh-Day et al, in prep.) in large datasets such as the Galaxy and Mass Assembly (GAMA) Survey (\citeauthor{2011MNRAS.413..971D} \citeyear{2011MNRAS.413..971D}, Liske et al. in prep).
\\
\noindent An initial application will involve using this method to make measurements of individual dark matter halo masses, and measurements of cosmological parameters, including $\sigma_{8}$, and $H_{0}$. \section*{Acknowledgments}

The referee is thanked for helping to clarify a number of issues in older versions, and contributing greatly to improving the quality of this paper. This research was conducted as part of the Australian Research Council Centre for Excellence for All-Sky Astrophysics (CAASTRO), through project number CE110001020. Special thanks goes to the members of the SAMI instrument and the SAMI survey teams for their advice, input and helpful discussions, in particular, Lisa Fogarty and Scott Croom. We acknowledge support from The University of Melbourne, and the Australian Astronomical Observatory (AAO).

\bibliography{bibliography}

\label{lastpage}
\end{document}

%% file: aasmacros.tex
\providecommand{\apjl}{ApJ}                
\providecommand{\apjs}{ApJS}               
\providecommand{\ao}{Appl.~Opt.}           
\providecommand{\aap}{A\&A}                

